\documentclass[aps,prd,reprint,preprintnumbers,showpacs,floatfix,nofootinbib,superscript address]{revtex4-2}

\usepackage[utf8]{inputenc}
\usepackage{parskip}
\usepackage{amssymb}
\usepackage{stix}
\usepackage{hhline}
\usepackage{amsmath}
\usepackage{mathtools}
\usepackage[dvipsnames]{xcolor}
\usepackage{xspace}
\usepackage{multirow,tabularx}
\usepackage{siunitx}
\usepackage{multirow}
\usepackage{graphicx}
\usepackage{xstring}
\usepackage{etoolbox}
\usepackage{notoccite}
\usepackage{natbib}
\usepackage{mathrsfs}
\usepackage{lineno}
\usepackage{tensor}
\usepackage{accents}
\usepackage{tikz}
\usepackage{listings}
\usepackage{epsfig,slashed,xcolor,bm,natbib,comment,enumitem,ulem,cancel}
\usepackage[cal=boondox]{mathalfa}

\allowdisplaybreaks

\def\vect#1{\mbox{\boldmath $#1$}}

\def\bg#1{{\mbox{\boldmath $ #1 $}}}
\def\pd#1#2{\frac{\partial #1}{\partial #2}}

\def\sspd#1#2{{\frac{\partial ^2 #1}{\partial #2^2}}}
\def\tfrac#1#2{{\textstyle\frac{#1}{#2}}}

\newcommand{\comma}{, }
\newcommand{\be}{\begin{equation}}
\newcommand{\ee}{\end{equation}}
\newcommand{\bea}{\begin{eqnarray}}
\newcommand{\eea}{\end{eqnarray}}
\newcommand{\myDel}[1]{{\color{red}\ifmmode\cancel{#1}\else\sout{#1}\fi}}

\newcommand{\kms}{{\rm\,km\,s^{-1}}}

\def\aj{\rm{AJ}}

\def\apj{\rm{ApJ}}
\def\apjl{\rm{ApJ}}
\def\apjs{\rm{ApJS}}

\def\aapr{\rm{A\&A~Rev.}}

\def\mnras{\rm{MNRAS}}

\def\pasj{\rm{PASJ}}

\def\astropj{\rm{Astrophys.~J.}}

\def\epjc{\rm{Eur.~Phys.~J.~C}}

\def\plettb{\rm{Phys.~Lett.~B}}

\def\pasj{\rm{Publ.~Astron.~Soc.~Jap.}}

\usepackage{graphicx}
\usepackage{dcolumn}
\usepackage{bm}

\usepackage{amsmath}

\usepackage{amssymb}

\usepackage{pgfplots}
\pgfplotsset{compat=newest}

\usepgfplotslibrary{external}
\usepackage{hyperref}

\hypersetup{
    colorlinks=true,
    linkcolor=blue,
    filecolor=red,      
    urlcolor=blue,
}

\usepackage{svg}
\svgpath{{../figures/}}

\usepackage{comment}

\newcommand{\barr}{\begin{array}}
\newcommand{\earr}{\end{array}}
\newcommand{\beqa}{\be \begin{array}{rcl}}
\newcommand{\eeqa}{\end{array} \ee}

\newcommand{\lam}{\lambda}

\newcounter{bean}
{\begin{list}%
{(\roman{bean})}{\topsep 0in \usecounter{bean}}}%
{\end{list}}

\graphicspath{{figures/}{extra_figures_for_notes/}}

\usepackage{hyperref}
\hypersetup{%
     colorlinks = true,%
     linkcolor = Blue,%
     citecolor = Blue,%
     filecolor = Blue,%
     urlcolor = Blue%
     }%
\usepackage[capitalize]{cleveref} 
\usepackage{siunitx}
\DeclareSIUnit\kiloparsec{kpc}

\begin{document}
\title{Gravitomagnetism and galaxy rotation curves: a cautionary tale} 
\author{A.N.~\surname{Lasenby}} \email{a.n.lasenby@mrao.cam.ac.uk}
\affiliation{Astrophysics Group, Cavendish Laboratory, JJ Thomson
  Avenue, Cambridge CB3 0HE\comma UK} \affiliation{Kavli Institute for
  Cosmology, Madingley Road, Cambridge CB3 0HA, UK}
\author{M.P.~\surname{Hobson}}
\email{mph@mrao.cam.ac.uk} \affiliation{Astrophysics Group, Cavendish
  Laboratory, JJ Thomson Avenue, Cambridge CB3 0HE\comma UK}
\author{W.E.V.~\surname{Barker}} \email{wb263@cam.ac.uk}
\affiliation{Astrophysics Group, Cavendish Laboratory, JJ Thomson
  Avenue, Cambridge CB3 0HE\comma UK} \affiliation{Kavli Institute for
  Cosmology, Madingley Road, Cambridge CB3 0HA, UK}

\begin{abstract}
We investigate recent claims that gravitomagnetic effects in linearised general relativity can explain flat and rising rotation curves, such as those observed in galaxies, without the need for dark matter. If one models a galaxy as an axisymmetric, stationary, rotating, non-relativistic and pressureless 'dust' of stars in the gravitoelectromagnetic (GEM) formalism, we show that gravitomagnetic effects on the circular velocity $v$ of a star are $O(10^{-6})$ smaller than the standard Newtonian (gravitoelectric) effects and thus any modification of galaxy rotation curves must be negligible, as might be expected. Moreover, we find that gravitomagnetic effects are $O(10^{-6})$ too small to provide the vertical support necessary to maintain the dynamical equilibrium assumed in such a model.  These issues are obscured if one constructs a single equation for $v$, as considered previously. We nevertheless solve this equation for a galaxy having a Miyamoto--Nagai density profile since this allows for both an exact numerical integration and an accurate analytic approximation.  We show that for the values of the mass, $M$, and semi-major and semi-minor axes, $a$ and $b$, typical for a dwarf galaxy, the rotation curve depends only very weakly on $M$, and becomes independent of it for larger $M$ values. Moreover, for aspect ratios $a/b > 2$, the rotation curves are concave over their entire range, which does not match observations in any galaxy. Most importantly, we show that for the poloidal gravitomagnetic flux $\psi$ to provide the necessary vertical support, it must become singular at the origin and have extremely large values near to it. This originates from the unwitting, but forbidden, inclusion of free-space solutions of the Poisson-like equation that determines $\psi$ and also clearly contradicts the linearised treatment implicit in the GEM formalism, hence ruling out the methodology in the form used as a means of explaining flat galaxy rotation curves. We further show that recent deliberate attempts to leverage such free-space solutions against the rotation curve problem yield no deterministic modification outside the thin disk approximation, and that, in any case, the homogeneous contributions to $\psi$ are ruled out by the boundary value problem posed by any physical axisymmetric galaxy.
\end{abstract}


\maketitle

\section{Introduction}
\label{sec:intro}

It is widely accepted that the modelling of galaxy rotation curves in
general relativity (GR) requires the inclusion of a dark matter halo
in order to reproduce observations
\cite{Lelli2016,Lelli2017,Li2020,Salucci2019}. In particular, the modelling of the
approximately flat rotation curves observed in the outskirts of large
spiral galaxies and, to a lesser extent, the rising rotation curves
observed in smaller dwarf galaxies
\cite{Rubin1970,Rubin1978,Rubin1980,Bosma1981,vanAlbada1985} is
considered to pose a significant challenge to GR without
such a component. The absence of any direct experimental evidence for
dark matter~\cite{Feng2010}  has thus led to the consideration of
various modified gravity theories to attempt to explain the
astrophysical data. 

There are a number of claims in the literature, however, that such
modifications are unnecessary since hitherto neglected effects in GR
itself are capable of explaining rotation curves without dark
matter. These include gravitoelectric flux confinement arising from
graviton self-interaction~\cite{Deur2009, Deur2014, Deur2017,
  Deur2019, Deur2020, Deur2021a, Deur2021b, Deur2022}, non-linear GR
effects arising even in the weak-gravity regime~\cite{Cooperstock2007}
and, most recently, gravitomagnetic effects in linearised GR
\cite{Ludwig2021}. Certain elements of~\cite{Ludwig2021} are further developed in~\cite{Astesiano:2022ghr} where (although the dark matter paradigm is not directly challenged) significant gravitomagnetic corrections to the rotation curve of a toy-model galactic baryon profile are suggested. An immediate question regarding such claims is how
such significant behaviours can have been consistently missed in the
long history of numerical relativity~\cite{Lehner2001, Lehner2014}, or
in the well-developed post-Newtonian formalism~\cite{Will2018a,
  Will2018b}. Perhaps unsurprisingly therefore, the claims in
\cite{Deur2009, Deur2014, Deur2017, Deur2019, Deur2020, Deur2021a,
  Deur2021b, Deur2022} and~\cite{Cooperstock2007} have been
subsequently shown to be non-viable in~\cite{Barker2022} and
\cite{Korzynski2007}, respectively. The purpose of this paper is to
perform the same function for the claims in~\cite{Ludwig2021,Astesiano:2022ghr}, by
showing that gravitomagnetism in the form used therein cannot be a significant factor in explaining 
flat or rising  galaxy rotation curves without dark matter. 

Our findings concur with the recent results reported in
\cite{Ciotti2022}, where the gravitoelectromagnetic (GEM) formulation of
linearised GR was used to predict galaxy rotation curves that at all
radii differ from those of Newtonian theory at the order of only
$v^2/c^2 \approx 10^{-6}$, as one might expect. The main focus of the
present paper, however, is to clarify {\it why} the approach adopted
in~\cite{Ludwig2021} leads to such different, unexpected and incorrect
results, which is not addressed in~\cite{Ciotti2022}. We will observe in particular the accidental involvement in~\cite{Ludwig2021} of homogeneous solutions to the GEM field equations: this leads us naturally to~\cite{Astesiano:2022ghr}, where such solutions are actively employed. We will show however that such solutions do not yield any deterministic phenomenology according to the suggested approximation in~\cite{Astesiano:2022ghr}, and that they are moreover absolutely ruled out by the absence of suitable GEM boundary conditions in the galactic environment.

The remainder of this paper is arranged as follows. In
Section~\ref{sec:lineargr}, we briefly outline linearised GR,
focussing on stationary non-relativistic matter sources, and discuss
its expression in the GEM formalism in Section~\ref{sec:gem}.  We then
summarise in Section~\ref{sec:gemmodel} the application of the GEM
formalism to the modelling of galaxy rotation curves, as proposed in
\cite{Ludwig2021}. We lay out the problems with this modelling
approach in Section~\ref{sec:problems}. Our analysis reveals the unwitting use of homogeneous solutions to the GEM field equations. In~\cref{HomogeneousPoloidalSolutions} we finally address the deliberate use in~\cite{Astesiano:2022ghr} of such solutions. Conclusions follow in
Section~\ref{sec:conc}.

\section{Linearised general relativity}
\label{sec:lineargr}

In the weak gravitational field limit appropriate for modelling galaxy
rotation curves, there exist quasi-Minkowskian coordinate systems
$x^\mu = (ct,x^i)$ in which the spacetime metric takes the form
$g_{\mu\nu} = \eta_{\mu\nu} + h_{\mu\nu}$ where $|h_{\mu\nu}| \ll 1$
and the first and higher partial derivatives of $h_{\mu\nu}$ are also
small.\footnote{We adopt the following sign conventions: $(+,-,-,-)$
  metric signature, ${R^\rho}_{\sigma\mu\nu} =
  2(\partial_{[\mu}{\Gamma^\rho}_{|\sigma|\nu]}
  +{\Gamma^\rho}_{\lambda[\mu}{\Gamma^\lambda}_{|\sigma|\nu]})$, where
  the metric (Christoffel) connection
  ${\Gamma^\rho}_{\lambda\mu}=\tfrac{1}{2}g^{\rho\sigma}(\partial_\lambda
  g_{\mu\sigma}+\partial_\mu g_{\lambda\sigma}-\partial_\sigma
  g_{\lambda\mu})$, and ${R^\rho}_{\mu} =
  {R^{\rho\sigma}}_{\mu\sigma}$. 
}. One can conveniently reinterpret $h_{\mu\nu}$ simply as a
special-relativistic symmetric rank-2 tensor field that represents the
weak gravitational field on a Minkowski background spacetime and
possesses the gauge freedom $h_{\mu\nu} \to h_{\mu\nu}
-\partial_\mu\xi_\nu - \partial_\nu\xi_\mu$.  Imposing the Lorenz
gauge condition $\partial_\rho \bar{h}^{\mu\rho}=0$ on the
trace-reverse $\bar{h}_{\mu\nu} \equiv
h_{\mu\nu}-\tfrac{1}{2}\eta_{\mu\nu} h$, where $h =
\eta_{\mu\nu}h^{\mu\nu}$, the linearised GR field equations reduce to
the simple form 
\be
\square^2 \bar{h}^{\mu\nu} = -2\kappa T^{\mu\nu},
\label{eqn:lineargrfe}
\ee
where $\square^2 \equiv \eta^{\mu\nu}\partial_\mu\partial_\nu$ is the
d'Alembertian operator, $\kappa = 8\pi G/c^4$ is Einstein's
gravitational constant and $T^{\mu\nu}$ is the matter energy-momentum
tensor.

For modelling galaxy rotation curves, it is sufficient to a very good
approximation to limit one's considerations to stationary,
non-relativistic, perfect fluid matter sources. In this case,
$\partial_0 T^{\mu\nu}=0$ and the coordinate 3-speed $u$ of any
constituent particle is small enough compared with $c$ that one may
neglect terms of order $u^2/c^2$ and higher in $T^{\mu\nu}$; in
particular one may take $\gamma_u = (1-u^2/c^2)^{-1/2} \approx
1$. Moreover, the fluid pressure $p$ is everywhere much smaller than
the energy density and may thus be neglected as a source for the
gravitational field.  Finally, we note that $|T^{ij}|/|T^{00}| \sim
u^2/c^2$ and so one should take $T^{ij} \approx 0$ to the order of our
approximation. Thus, for a stationary, non-relativistic source, one
approximates its energy-momentum tensor as
\be
T^{00} \approx \rho c^2,\qquad T^{i0} \approx
c\rho u^i, \qquad T^{ij} \approx 0,
\label{eqn:emtensor}
\ee
where $\rho(\vect{x})$ is the proper-density distribution of the
source and $\vect{x}$ denotes a spatial 3-vector.  As an immediate
consequence, the particular integral of (\ref{eqn:lineargrfe}) yields
$\bar{h}^{ij} \approx 0$. Indeed, this is consistent with the Lorenz
gauge condition, which implies that
$\partial_j\bar{h}^{ij}=-\partial_0\bar{h}^{i0}$, where the right-hand
side vanishes for stationary systems. Thus, only the $\bar{h}^{00}$
and $\bar{h}^{0i} = \bar{h}^{i0}$ components of the gravitational field
tensor are non-zero in this approximation.

In linearised GR, there is an inconsistency between the field
equations (\ref{eqn:lineargrfe}) and the equations of motion for
matter in a gravitational field. From (\ref{eqn:lineargrfe}), one
quickly finds that $\partial_\mu T^{\mu\nu} = 0$, which should be
contrasted with the requirement from the full GR field equations that
the covariant divergence should vanish, $\nabla_\mu T^{\mu\nu}=0$.
The latter requirement leads directly to the geodesic equation of
motion for the worldline $x^\mu(\tau)$ of a test particle, namely
\be
\ddot{x}^\mu + {\Gamma^\mu}_{\nu\sigma}\dot{x}^\nu\dot{x}^\sigma
= 0,
\label{eqn:geodesic}
\ee
where the dots denote differentiation with respect to the proper time
$\tau$, whereas the former requirement leads to the equation of motion
$\ddot{x}^\mu = 0$. This means that the gravitational field has {\it
  no effect} on the motion of the particle and so clearly contradicts
the geodesic postulate. Despite this inconsistency, one may show that
the effect of weak gravitational fields on test particles may still be
computed by inserting the linearised connection coefficients into the
geodesic equations (\ref{eqn:geodesic}).

\section{Gravitoelectromagnetism}
\label{sec:gem}

Gravitoelectromagnetism (GEM) provides a useful and
notionally-familiar formalism for linearised GR by drawing a close
analogy with classical electromagnetism (EM). Indeed, GEM is ideally
suited to modelling galaxy rotation curves, since the assumption of
a stationary, non-relativistic matter source leads to GEM field
equations and a GEM `Lorentz' force law (derived below) that are fully
consistent and have forms analogous to their counterparts in EM; this
is not possible for more general time-dependent scenarios.

The GEM formalism for linear GR with a stationary, non-relativistic
source is based on the simple ansatz of relabelling\footnote{Conventions in the literature vary up to a multiplicative
  constant for the definition of the gravitomagnetic vector potential
  $A^i$. These factors variously modify the analogues of the EM field
  equations and the Lorentz force law, with no scaling choice allowing
  all the GEM and EM equations to be perfectly analogous. Here, we
  follow the convention used in~\cite{GRbook}.} the four
independent non-zero components of $\bar{h}^{\mu\nu}$
as $\bar{h}^{00}\equiv
4\Phi/c^2$ and $\bar{h}^{0i} \equiv A^i/c$, where we have
defined the gravitational scalar potential $\Phi$ and spatial
gravitomagnetic vector potential $A^i$. On lowering indices, the
corresponding components of $h_{\mu\nu}$ are $h_{00}=
h_{11}=h_{22}=h_{33}=2\Phi/c^2$ and $h_{0i}=A_i/c$.  It should be
remembered that raising or lowering a spatial (Roman) index introduces
a minus sign with our adopted metric signature. Thus the numerical
value of $A_i$ is minus that of $A^i$, the latter being the $i$th
component of the spatial vector $\vect{A}$.  It is also worth noting
that both $\Phi/c^2$ and $A_i/c$ are dimensionless, thereby yielding
dimensionless components $h_{\mu\nu}$, which is consistent with our
choice of coordinates $x^\mu = (ct,x^i)$ having dimensions of length.

With the above identifications, the linearised field equations
(\ref{eqn:lineargrfe}) with energy-momentum tensor (\ref{eqn:emtensor})
may be written in the scalar/vector form
\be
\nabla^2 \Phi  = 4\pi G \rho, \qquad 
\nabla^2 \vect{A}  = \frac{16 \pi G}{c^2} \vect{j},
\label{eqn:gemfe}
\ee
where we have defined the momentum density (or matter current density)
$\vect{j}\equiv \rho\vect{u}$, and the Lorenz gauge condition
$\partial_\rho \bar{h}^{\mu\rho}=0$ itself becomes
$\bg{\nabla}\cdot\vect{A}=0$.  Clearly, the first equation in
(\ref{eqn:gemfe}) recovers the Poisson equation for the gravitational
potential, familiar from Newtonian gravity, whereas the second
equation determines the gravitomagnetic vector potential that
describes the `extra' (weak) gravitational field predicted in
linearised GR, which is produced by the motion of the fluid
elements in a stationary, non-relativistic source.  Indeed, the
general solutions to the equations (\ref{eqn:gemfe}) are given
immediately by
\begin{subequations}
\label{eqn:gemsolutions}
\bea
\Phi(\vect{x}) & = & -G \int \frac{\rho(\vect{x}')}{|\vect{x}-
	\vect{x}'|}\,\mathrm{d}^3\vect{x}', \label{eqn:gemsolutionsPhi}\\
\vect{A}(\vect{x}) & = & -\frac{4G}{c^2} \int
	\frac{\vect{j}(\vect{x}')}{|\vect{x}-\vect{x}'|}\,\mathrm{d}^3\vect{x}'.
\label{eqn:gemsolutionsA}
\eea
\end{subequations}

One may take the analogy between linearised GR and EM further by
defining the gravitoelectric and gravitomagnetic fields
$\vect{E} = -\vect{\nabla}\Phi$ and $\vect{B} =
\vect{\nabla}\times\vect{A}$, which are easily found to
satisfy the gravitational Maxwell equations
\be
\begin{array}{rl@{\qquad}rl}
\vect{\nabla} \cdot \vect{E} &= -4\pi
G\rho, & \vect{\nabla} \cdot \vect{B} &= 0, \\[6pt]
\vect{\nabla} \times\vect{E} &= \vect{0}, & \vect{\nabla}
\times\vect{B}  &= -\dfrac{16\pi G}{c^2} \vect{j}.
\end{array}
\ee
The gravitoelectric field $\vect{E}$ describes the standard
(Newtonian) gravitational field produced by a static matter
distribution, whereas the gravitomagnetic field $\vect{B}$ is the
`extra' gravitational field produced by moving fluid elements in
the stationary, non-relativistic source.

The equation of motion for a test particle in the presence of the
GEM fields is merely the geodesic equation
(\ref{eqn:geodesic}) for the metric $g_{\mu\nu} = \eta_{\mu\nu} +
h_{\mu\nu}$, from which one may determine the trajectories of either
massive particles, irrespective of their speed, or massless particles,
by considering timelike or null geodesics, respectively.  We will
assume here, however, that the test particle is massive and
slowly-moving, i.e. its coordinate 3-speed $v$ is sufficiently small
that we may neglect terms in $v^2/c^2$ and higher. Hence we may take
$\gamma_v = (1-v^2/c^2)^{-1/2} \approx 1$, so that the 4-velocity of
the particle may be written $v^\mu = \gamma_v(c,\vect{v}) \approx
(c,\vect{v})$.  This immediately implies that $\ddot{x}^0=0$ and,
moreover, that $\mathrm{d}t/\mathrm{d}\tau = 1$, so one may consider only the spatial
components of (\ref{eqn:geodesic}) and replace dots with derivatives
with respect to $t$. Expanding the summation in (\ref{eqn:geodesic})
into terms containing, respectively, two time components, one time and
one spatial component, and two spatial components, neglectng the
purely spatial terms since their ratio with respect to the purely
temporal term is of order $v^2/c^2$, expanding the connection
coefficients to first-order in $h_{\mu\nu}$ and remembering that for a
stationary field $\partial_0 h_{\mu\nu}=0$ and that one inherits a
minus sign on raising or lower a spatial (Roman) index, one finally
obtains the gravitational Lorentz force law
\be
\frac{\mathrm{d}\vect{v}}{\mathrm{d}t} = -\vect{\nabla}\Phi +
\vect{v}\times(\vect{\nabla}\times\vect{A}) = 
\vect{E} +
\vect{v}\times\vect{B}.
\label{eqn:gemeom}
\ee
The first term on the right-hand side gives
the standard Newtonian result for the motion of a test particle in
the field of a static, non-relativistic source, whereas the second
term gives the `extra' force felt
by a moving test particle in the presence of the `extra' field
produced by moving fluid elements in the stationary, non-relativistic
source.

\section{Gravitoelectromagnetic modelling of galaxy rotation curves}
\label{sec:gemmodel}

The GEM formalism is applied to the modelling of galaxy rotation
curves in~\cite{Ludwig2021}, where the galactic density and velocity
distribution is assumed to act as a stationary, non-relativistic
matter source. Thus, somewhat unusually, the fluid pressure is assumed
to vanish and the galaxy is instead modelled as consisting of a `dust' of
stars. This approach therefore uses the field equations
(\ref{eqn:gemfe}) and the equation of motion (\ref{eqn:gemeom}), where
the velocity distribution $\vect{u}$ of the galaxy in the former is
identified with the velocity $\vect{v}$ of test particles in the
latter, thereby leading to a potentially self-consistent pressureless model.

The central result in~\cite{Ludwig2021} can be derived straightforwardly as
follows. First, one adopts cylindrical polar coordinates $(R,\phi,z)$
and assumes azimuthmal symmetry, such that $\rho=\rho(R,z)$ and
$\vect{v} = v(R,z)\hat{\bg{\phi}}$, which from
(\ref{eqn:gemsolutions}) implies that
$\Phi = \Phi(R,z)$ and $\vect{A} = A(R,z)\hat{\bg{\phi}}$. In this
case,
\begin{subequations}
\bea
\vect{\nabla}\times\vect{A} &=&
\frac{1}{R}\left(-\pd{\psi}{z}\hat{\vect{R}} 
+ \pd{\psi}{R}\hat{\vect{z}}\right),\\
\vect{v}\times(\vect{\nabla}\times\vect{A}) &=&
\frac{v}{R}\left(\pd{\psi}{R}\hat{\vect{R}} + \pd{\psi}{z}\hat{\vect{z}}\right),
\eea
\end{subequations}
where we have defined the poloidal gravitomagnetic flux $\psi \equiv
RA$.  Also, in light of the Lorenz (or Coulomb) gauge condition
$\vect{\nabla}\cdot\vect{A}=0$ (which is easily confirmed by direct
calculation), one has
\be \nabla^2\vect{A} =
-\vect{\nabla}\times(\vect{\nabla}\times\vect{A}) =
\left[\pd{}{R}\left(\frac{1}{R}\pd{\psi}{R}\right) +
  \frac{1}{R}\sspd{\psi}{z}\right]\hat{\vect{\phi}}.  
\ee
The field equations (\ref{eqn:gemfe}) and the radial and vertical
components of the fluid equation of motion (\ref{eqn:gemeom}) may
therefore be written as
\begin{subequations}
\label{eqn:gemsystem}
\bea
\frac{1}{R}\pd{}{R}\left(R\pd{\Phi}{R}\right) + \sspd{\Phi}{z} & = &
4\pi G \rho, \label{poissonphi}\\
\pd{}{R}\left(\frac{1}{R}\pd{\psi}{R}\right) +
\frac{1}{R}\sspd{\psi}{z} & = & \frac{16\pi G}{c^2}\rho v,\label{poissona}\\
\pd{\Phi}{R}-\frac{v}{R}\pd{\psi}{R} & = & \frac{v^2}{R}\label{eomrad},\\
-\pd{\Phi}{z} + \frac{v}{R}\pd{\psi}{z} & = & 0.\label{eomazi}
\eea
\end{subequations}
Using (\ref{eomrad}) and (\ref{eomazi}) to eliminate
$\partial\psi/\partial R$ and $\partial\psi/\partial z$ from
(\ref{poissona}), then using (\ref{poissonphi}) to eliminate the
resulting term containing $\partial^2\Phi/\partial z^2$, the field
equation (\ref{poissona}) yields
\bea
\hspace*{-1cm}\left(v+R\pd{v}{R}\right)\pd{\Phi}{R} +
  R\pd{v}{z}\pd{\Phi}{z} = && \nonumber \\
&& \hspace*{-3.5cm} \frac{v}{R}\left[v\left(v -
    R\pd{v}{R}\right)+4\pi G\rho
    R^2\left(1-\frac{4v^2}{c^2}\right)\right].
\label{ludwigeqn}
\eea

The non-linear first-order partial differential equation
(\ref{ludwigeqn}) for the galactic velocity field $v(R,z)$ is the key
expression in~\cite{Ludwig2021}\footnote{Equation (\ref{ludwigeqn})
  does, in fact, differ slightly from equation (4.1) in
 ~\cite{Ludwig2021}, since the latter lacks the factor of $4$
  muliplying $v^2/c^2$ in the final term on the RHS. We believe the
  expression in~\cite{Ludwig2021} to be in error as a consequence of
  the choice of scaling used in the definition therein of the
  gravitomagnetic vector potential $\vect{A}$.}, and depends only on
the galactic density distribution $\rho$ and on the derivatives
$\partial\Phi/\partial R$ and $\partial\Phi/\partial z$ of the
Newtonian gravitational potential, which are themselves also
determined by specifying $\rho$. Indeed, $\Phi$ is given by
(\ref{eqn:gemsolutionsPhi}), which in cylindrical polar coordinates
with azimuthal symmetry reads\footnote{Our final expression for $\Phi$
  differs by a factor of 2 as compared to equation (4.2) in [1]; we
  believe the latter to be in error.}
  \begin{equation}
  \begin{aligned}
\hspace*{-0.6cm}\Phi(R,z) &= -G\!\!\int_0^\infty \!\!\! \mathrm{d}R' \!\!\int_0^{2\pi} \!\!\! \mathrm{d}\phi'
\!\!\int_{-\infty}^\infty \!\!\! \mathrm{d}z'\, 
\frac{R'\rho\left(R',z'\right)}{|\vect{x}-\vect{x}'|} \\ 
	  &=
-2G\!\! \int_0^\infty \!\!\! \mathrm{d}R' \!\! \int_{-\infty}^\infty \!\!\!\! \mathrm{d}z'
\rho\left(R',z'\right) R'\sqrt{\frac{m}{RR'}}K(m),
\label{phisol}
  \end{aligned}
  \end{equation}
where $K(m)$ is a complete elliptic integral function of the first
kind and $m = 4RR'/[(R+R')^2 + (z-z')^2]$. Moreover, the 
derivatives $\partial\Phi/\partial R$ and $\partial\Phi/\partial z$
may also be expressed analytically as
  \vspace{5pt}
\begin{widetext}
\begin{subequations}
	\bea \pd{\Phi}{R} & = & G\int_0^\infty \mathrm{d}R' \int_{-\infty}^\infty \mathrm{d}z'\,
\rho\left(R',z'\right) \frac{R'}{R}\sqrt{\frac{m}{RR'}}\left[ K(m) +
  \tfrac{1}{2}\left(\frac{R}{R'}-\frac{2-m}{m}
  \right)\frac{mE(m)}{1-m}\right], \label{dphidr} \\ \pd{\Phi}{z} & =
	& \frac{G}{2}\int_0^\infty \mathrm{d}R' \int_{-\infty}^\infty \mathrm{d}z'\,
\rho\left(R',z'\right)\left(\frac{z-z'}{R}\right)\sqrt{\frac{m}{RR'}}\frac{mE(m)}{1-m},\label{dphidz}
\eea
\end{subequations}
\end{widetext}
where $E(m)$ denotes a complete elliptic integral of the second kind.

Before considering further the application of equation
(\ref{ludwigeqn}) to modelling galaxy rotation curves, we note that,
if one neglects the mass currents on the RHS of (\ref{poissona}) (by
letting $c \to \infty$), then one may consistently set $\psi = 0$
(although other solutions to the resulting homogeneous equation
(\ref{poissona}) do exist). The radial and vertical components of the
fluid equation of motion (\ref{eomrad})--(\ref{eomazi}) then immediately
yield $\partial\Phi/\partial z = 0$ and thus $v^2(R) =
R\,\partial\Phi/\partial R$, where the latter is the usual Newtonian
equation assumed in the modelling of galaxy rotation curves.

In applying the full equation (\ref{ludwigeqn}) to the modelling of
galaxy rotation curves, it is noted in~\cite{Ludwig2021} that observations
of the rotation velocity are typically made along the galactic
equatorial plane, so one may take $z=0$. Assuming further a galactic density
distribution that is symmetric about this mid-plane, (\ref{ludwigeqn})
then reduces to
\bea
\hspace*{-0.7cm}\left(\beta+R\pd{\beta}{R}\right)\pd{\Phi(R,0)}{R} = && \nonumber \\
&&\hspace*{-3.5cm}\frac{c^2\beta}{R}\left[\beta\left(\beta - R\pd{\beta}{R}\right)+\frac{4\pi G}{c^2}\rho(R,0)
  R^2(1-4\beta^2)\right],
\label{ludwigeqn2}
\eea
where we have defined $\beta(R) \equiv v(R,0)/c$. Equation
(\ref{ludwigeqn2}) is applied in~\cite{Ludwig2021} to two different models of
the galactic density distribution.

The first model considered uses the density and gravitational
potential given by the analytical Miyamoto--Nagai (MN) solution to
Poisson's equation~\cite{Miyamoto1975}. In this approach, one begins
by assuming the fairly simple potential form
\be
\Phi(R,z)  =  -\frac{GM}{\sqrt{R^2+(a+\sqrt{b^2+z^2})^2}},
\label{eq:MNPhi}
\ee
where $M$ is the total galactic mass and $a$ and $b$ are free positive
parameters. The density distribution implied by Poisson's equation is
then given by
\be
\begin{aligned}
	\rho(R,z) & =  \\
	&\hspace{-15pt}\frac{Mb^2}{4\pi}\times
\frac{aR^2 + (a + 3\sqrt{b^2+z^2})(a+\sqrt{b^2+z^2})^2}
{\left[R^2+(a+\sqrt{b^2+z^2})^2\right]^{5/2}(b^2+z^2)^{3/2}},
\label{eq:MNrho}
\end{aligned}
\ee
which extends to infinity in both $R$ and $z$. The constant density
contours have the form of spheroids of revolution with semi-axes
proportional to $a$ and $b$. It is straightforward to verify that,
when integrated over all space, this density distribution yields the
total mass $M$. In~\cite{Ludwig2021}, this model is fitted to the observed
rising rotation curve of NGC 1560 out to $\SI{8.3}{\kiloparsec}$ by varying the
parameters $M$, $a$ and $b$. The derived parameter values are $M = 7.3
\times 10^{10}$ M$_\odot$, $a =\SI{0.373}{\kiloparsec}$ and $b =\SI{0.300}{\kiloparsec} $, which
yield a reasonable fit to the rotation curve, but does not reproduce
the luminosity profile of NGC 1560. This occurs because the infinite
spheroidal solution does not describe the equilibrium of a finite
disk-like object, and thus fails to reproduce its mass distribution
and total mass.

Consequently, in the second model, the galaxy is instead considered as an
axisymmetric thin disk of finite radius, which is again symmetric
about its mid-plane $z=0$. The density distribution is assumed to have
the functional form
\be
\rho(R,z) = \rho(R,0)\exp\left(-\frac{z^2}{2\Delta^2(R)}\right),
\label{thindisk}
\ee
where $\Delta(R)$ is a characteristic disk width with some assumed
radial dependence. For small values of $\Delta(R)$, one can estimate
the integral over $z'$ in (\ref{dphidr}) analytically using the
Laplace approximation, which boils down to setting $z'=0$ in the
integrand and multiplying by the volume $\sqrt{2\pi}\Delta(R)$ of the
Gaussian factor in (\ref{thindisk}); this yields
\begin{widetext}
\be
\pd{\Phi(R,0)}{R}  \approx  2\sqrt{2\pi}G\int_0^\infty
\frac{R'\rho(R',0)\Delta(R')}{R(R+R')}\left[K\left(\frac{4RR'}{(R+R')^2}\right)
	+ \frac{R+R'}{R-R'}E\left(\frac{4RR'}{(R+R')^2}\right)\right]\,\mathrm{d}R'.
\ee
\end{widetext}
To evaluate the above integral (numerically), the density distribution
$\rho(R,0)$ is taken from the luminosity profile of the galaxy under
consideration, which is therefore reproduced {\it automatically}, but
one still requires a model for the radially-dependent characteristic
vertical width $\Delta(R)$ of the galaxy. In~\cite{Ludwig2021}, this is taken to
coincide with a given constant density contour of the analytical
MN solution (\ref{eq:MNrho}). In particular, one defines
$\Delta(R)$ such that
\be
\frac{\rho_{MN}(R,\Delta(R))}{\rho_{MN}(0,0)}\frac{{\cal M}(\ell)}{M} = \exp\left(-\frac{\ell^2}{2}\right), 
\label{eqn:delta}
\ee
where $\rho_{MN}(R,z)$ denotes the right-hand side of
(\ref{eq:MNrho}), $\ell$ is a pre-defined `label', which is usually
set to $\ell = 3$ so that the chosen contour contains a fraction
(approximated by $1 -
\exp(-\ell^2/2)=0.989$) of the total mass $M$, and ${\cal M}(\ell)$ is
the resulting approximate mass of the galaxy with the density
distributon (\ref{thindisk}), estimated by using the Laplace approximation to
perform the integral over $z$:
\be
{\cal M}(\ell) = (2\pi)^{3/2}\int_0^{R_{\rm max}} \Delta\left(R'\right)\rho\left(R',0\right)R'\,\mathrm{d}R',
\ee
where the maximum radius of the galactic disk, $R_{\rm max}$, is
obtained by solving (\ref{eqn:delta}) with $\Delta(R) = 0$. This
approach requires quite a time-consuming iterative process, but the
velocity profile again depends only on the three free parameters $M$,
$a$ and $b$. When fitted to the same observed rising rotation curve
data for NGC 1560 as used above, the derived parameter values for this
model are $M = 1.52 \times 10^{10}$ M$_\odot$, $a = \SI{7.19}{\kiloparsec}$ and $b =
\SI{0.567}{\kiloparsec}$ (yielding $R_{\rm max} = \SI{12.2}{\kiloparsec}$), which again produces a
reasonable fit to the rotation curve, but now also reproduces the
luminosity profile of NGC 1560 by construction. The model is also used
in~\cite{Ludwig2021} to reproduce satisfactorily the observed rotation curve
data for the spiral galaxy NGC 3198 and the lenticular galaxy NGC 3115.

\section{Problems with the model}
\label{sec:problems}

Although the approach outlined above appears at first sight to be a
reasonable methodology for modelling galaxy rotation curves using the
GEM formalism, it does have some unusual features. As mentioned
previously, the model most notably assumes that the galaxy consists of
a pressureless `dust' of stars, all of which follow circular
orbits. In particular, this means that the vertical support necessary
to maintain dynamical equilibrium is assumed all to arise from
gravitomagnetic rotational effects, which we will see leads to such
effects being massively overestimated. This shortcoming may be
addressed by, for example, using a distribution function approach
based on the GEM formulation of the Jeans equation~\cite{Ciotti2022},
since this enables vertical support via a velocity dispersion of the
stars, and also allows for individual stars to follow non-circular
orbits whilst retaining net currents that are strictly azimuthal.

Whilst the more sophisticated approach of distribution functions makes
better physical sense, we will not concern ourselves with such
modifications here, since we wish merely to address why the
methodology outlined in Section~\ref{sec:gemmodel} can lead to
incorrect conclusions regarding the effect of gravitomagnetism on
galactic rotation curves.  Indeed, we will limit our considerations
still further by choosing not to pursue the iterative numerical
process for obtaining rotation curves for the thin disk density model
(\ref{thindisk}), since this is rather computationally cumbersome and
time consuming. Instead, we will restrict our attention here to the
model having the MN density profile (\ref{eq:MNrho}), which can be
treated almost entirely analytically and suffices to demomstrate the
shortcomings of the overall approach outlined in
Section~\ref{sec:gemmodel}.

\subsection{Order of magnitude analysis}

Before considering the single key equation (\ref{ludwigeqn}) that
forms the basis of the approach outlined in
Section~\ref{sec:gemmodel}, we begin by making some observations
regarding the four separate equations (\ref{eqn:gemsystem})
from which (\ref{ludwigeqn}) is derived. In particular, we first note
that in the radial equation of motion (\ref{eomrad}) one requires a
term in $v^2$ to obtain sensible results. This occurs because $v$ is
${\cal O}(\sqrt{\Phi})$ rather than ${\cal O}(\Phi)$.
Indeed, one can see from the set of equations
(\ref{eqn:gemsystem}) that in typical circumstances there
will exist a hierarchy of magnitudes for different quantities, and it
is worth describing this hierarchy now so as to orient ourselves.

This is most easily achieved by first adopting geometric units $G = c
= 1$, which we will assume henceforth. One then requires only a single
scale to specify the base of units, which in this application is most
conveniently taken to be length. In particular, we take the unit of
length to be $\SI{1}{\kiloparsec}$, which corresponds to typical galactic scales. All
other physical quantities can then be expressed in terms of this base
unit. For example, a mass $M$ in SI units is given in terms of our
units by $GM/c^2\times\SI{1}{\kiloparsec\per\meter}$, whereas a density $\rho$
  in SI units is given by $G\rho/c^2 \times \SI{1}{\kiloparsec\squared\per\meter\squared} \approx
  7.07 \times 10^{11} \rho$. Thus, typical galactic densities of $\sim
  \SI{2e-19}{\kilogram\per\meter\cubed}$ correspond to $\sim 1.4 \times
  10^{-7}$ in our units. For a broad selection of galaxies types, one
  may therefore take typical densities in our units to lie in the
  range ${\cal O}(10^{-8})$ to ${\cal O}(10^{-6})$; we will take the
  upper of these as indicative, since this maximises the magnitude of
  gravitomagnetic effects, although in reality they will usually be
  somewhat smaller.

\begin{figure*}
\begin{center}
\includegraphics[width=0.99\textwidth]{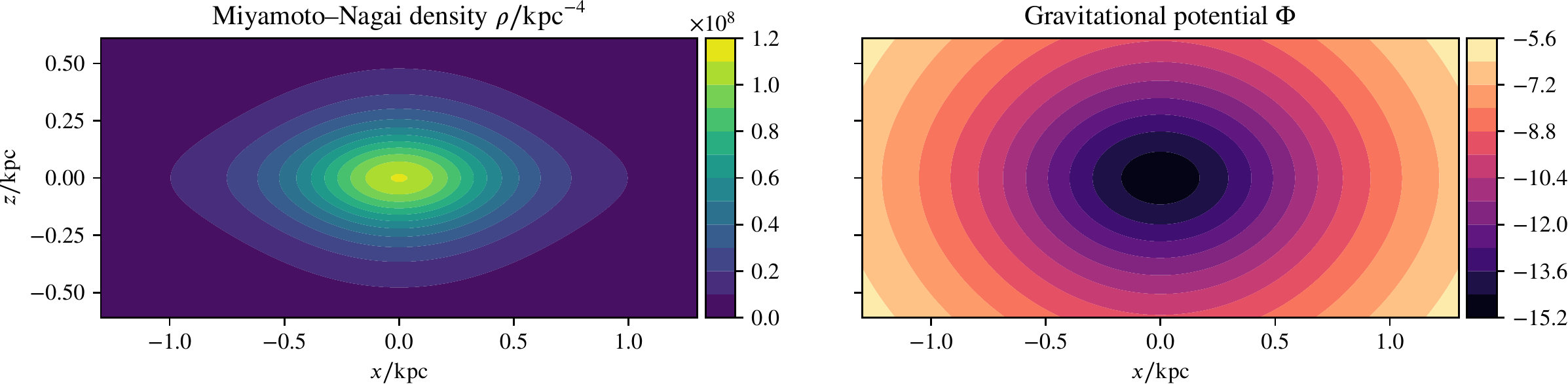}
\caption{Gravitational density (left) and potential (right) contours
  for a MN profile with parameters $a$ and $b$ derived from fitting
  the rotation curve of NGC 1560 in~\cite{Ludwig2021}. \label{fig:ngc1560-contours}}
\end{center}
\end{figure*}

From the Poisson equation (\ref{poissonphi}), or its more succinct
form in (\ref{eqn:gemfe}), one sees that $|\Phi|$ is also ${\cal
  O}(10^{-6})$, and hence the velocity $v \sim {\cal O}(10^{-3})$
(where to convert velocities in SI unit to our units, one
needs merely to divide by $c$).
Then, from equation (\ref{poissona}), which one can also write more
usefully as $\nabla^2 \psi = 16\pi R\rho v -
(2/R)\partial\psi/\partial R$, one sees that $|\psi| \sim {\cal
  O}(10^{-9})$, modulo any multiplicative effects from $R$ which
are limited to a factor of $\sim 10$ for a typical galaxy.

Now considering either the radial equation of motion (\ref{eomrad}) or
its vertical counterpart (\ref{eomazi}), one sees any effects arising
from $\psi$, which always appears multiplied by $v$, must be ${\cal
  O}(10^{-6})$ {\it smaller} that those arising from
$\Phi$. Consequently, any gravitomagnetic effects will have a negligible
effect on the circular velocity of a test particle, which will very
well approximated simply by the strictly Newtonian expression
$\sqrt{R\,\partial\Phi/\partial R}$.

This result is at least allowable (notwithstanding the usual
clash, if no dark matter is assumed, with the flat or rising rotation curves observed in many
galaxies), if disappointing, but one sees from the vertical equation
of motion (\ref{eomazi}) that there is a much more serious problem.
In this case, one requires the ${\cal O}(10^{-6})$ term in $\Phi$ to
be balanced by the ${\cal O}(10^{-12})$ term in $\psi$; this is simply
impossible and indicates that the set of equations
(\ref{eqn:gemsystem}) has no physically meaningful
solution. As mentioned above, this problem arises because one has
insisted that all the vertical support force arises from
gravitomagnetic effects, which is impossible for ordinary matter.

\subsection{Rotation curves for the MN density profile}

In eliminating various quantities between the equations
(\ref{eqn:gemsystem}) to arrive at the `master' equation
(\ref{ludwigeqn}) in~\cite{Ludwig2021}, one can no longer identify the issues
discussed above. Indeed, one can go on to find solutions $\psi$ that
satisfy (\ref{ludwigeqn}), although these cannot be physically
meaningful, as our analysis above shows. We now illustrate this
directly by considering a galaxy having the MN density profile
(\ref{eq:MNrho}) and gravitational potential (\ref{eq:MNPhi}), which
was the first model used in~\cite{Ludwig2021} to fit the observed rotation curve
data of NGC 1560 (although it fails to reproduce its luminosity
profile).  As discussed above, the resulting derived parameters are $M
= 7.3 \times 10^{10}$ M$_\odot$, $a = \SI{0.373}{\kiloparsec}$ and $b = \SI{0.300}{\kiloparsec}$,
so the fitted MN density profile is moderately oblate. The resulting
gravitational potential and density contours are shown in
Figure~\ref{fig:ngc1560-contours}.

Inserting the forms for the MN potential (\ref{eq:MNPhi}) and density
(\ref{eq:MNrho}) into the `master' equation (\ref{ludwigeqn}) yields a
very complicated expression, but one can make progress analytically if
one restricts attention to the equatorial plane $z=0$, as in
(\ref{ludwigeqn2}). This is permissible since, although
(\ref{ludwigeqn}) contains the $z$-derivative of the potential
$\Phi$, one can see that for the MN form of the potential this vanishes on
the equatorial plane. The resulting equation then reads $A+B = 0$, with
\begin{widetext}
\begin{subequations}
\label{eqn:radial-chi-eqn}
\bea
A &=& -R^{2} M\left[-b R\left(R^{2}+a^{2}+2 b a+b^{2}\right)\frac{dv}{d R}
+\left(2 b^{3}+5 b^{2} a+\left(4 a^{2}-R^2\right) b+R^{2}
a+a^{3}\right)v\right] \label{eqn:radial-chi-eqnA} \\
B & = & 4\left[\tfrac{1}{4} b\left(R^{2}+a^{2}+2 b a+b^{2}\right)^{5/2}\left(\frac{dv}{d R} R-v\right)+M R^{2} v\left(5 a^{2} b+a^{3}+R^{2} a+3 b^{3}+7 b^{2} a\right)\right] v^{2},
\label{eqn:radial-chi-eqnB}
\eea
\end{subequations}
\end{widetext}
where we have split the LHS into the terms, since it is possible to
obtain a simple analytic result for $v$ by just setting $A = 0$.  It
is not immediately obvious that this is a valid procedure, even as an
approximation, since $v \sim {\cal O}(10^{-3})$ and $\rho$, and by
extension its volume integral $M$, are likely ${\cal
  O}(10^{-6})$. Thus, both expression $A$ and the first half
of the terms in $B$ are likely ${\cal O}(10^{-9})$, and hence it is not
clear that one can preferentially drop the first half of $B$.
Numerically, however, it transpires that the value of $M=7.3 \times
10^{10}$~M$_\odot$ derived for NGC 1560 is sufficiently large that one
can consider just $A=0$, and we note that this yields an
expression for $v$ that is in fact {\it independent} of $M$.

\begin{figure}
\begin{center}
\includegraphics[width=0.99\linewidth]{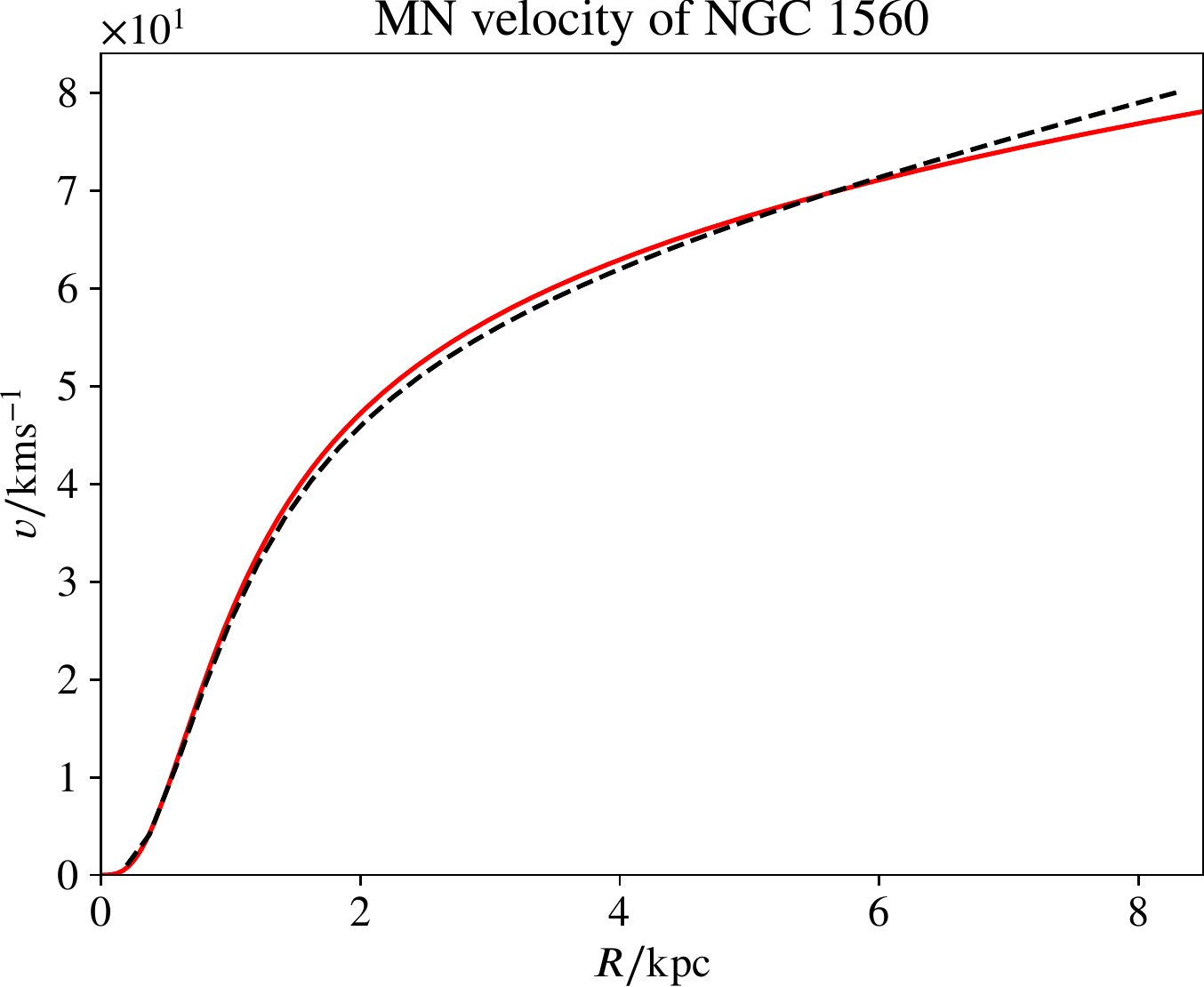}
\caption{Rotation velocity $v$ (in units of $c$) versus $R$ in $\SI{}{\kiloparsec}$ for
  a MN profile with parameters derived from NGC 1560. The red curve is
  obtained using the analytical approximation (\ref{eqn:anal-chi})
  with $C=1/6400$ and the black curve is an exact numerical
  integration using equation (\ref{eqn:radial-chi-eqn}).}
\label{fig:ngc-1560-rot-curves}
\end{center}
\end{figure}

We may illustrate this approach explicitly by comparing the exact and
approximate solutions for $v$ in this case. Setting just $A=0$ and
solving for $v$ gives
\be
v=\frac{C R^{2+\frac{a}{b}}}{[R^2+(a+b)^2]^{3/2}},
\label{eqn:anal-chi}
\ee
where $C$ is an arbitrary constant. In Figure~\ref{fig:ngc-1560-rot-curves},
we show the rotation curve resulting from the analytical approximation
(\ref{eqn:anal-chi}) as the red curve and an exact numerical
integration of the full equation (\ref{eqn:radial-chi-eqn}) as the
black curve. For the analytic approximation, although there is no
dependence on mass, one must provide an overall scaling $C$, and a
value of $C=1/6400$ was used in the plot, which gives reasonably good
agreement between the exact result in this case. The latter was
calculated by numerical integration starting at the outermost rotation
curve data point for NGC 1560, for which $v= 2.67 \times 10^{-4}$ (in
units of $c$) at $R=\SI{8.29}{\kiloparsec}$, and moving inwards towards the origin,
in the same way as performed in~\cite{Ludwig2021}. Similarly, one could instead
fix the scaling $C$ of the analytical result by ensuring that it
passes through the outermost data point, which moves the red curve up
slightly.

In any case, it is important to note that, while the fit to the NGC
1560 rotation curve data in~\cite{Ludwig2021} yields the derived mass $M=7.3
\times 10^{10}$~M$_\odot$, the only information about $M$ is in quite
small changes in the {\it shape} of the curve that occur as $M$ drops
below this best-fit value. For {\it larger} values of $M$, the shape
of the curve is invariant, and corresponds to that given in the
analytical approximation (\ref{eqn:anal-chi}), which does not depend
on $M$. This suggests that there may be a large uncertainty on the
mass $M$ derived from the rotation curve data, although no errors on
the fitted value are provided in~\cite{Ludwig2021}.

\begin{figure}
\begin{center}
\includegraphics[width=0.99\linewidth]{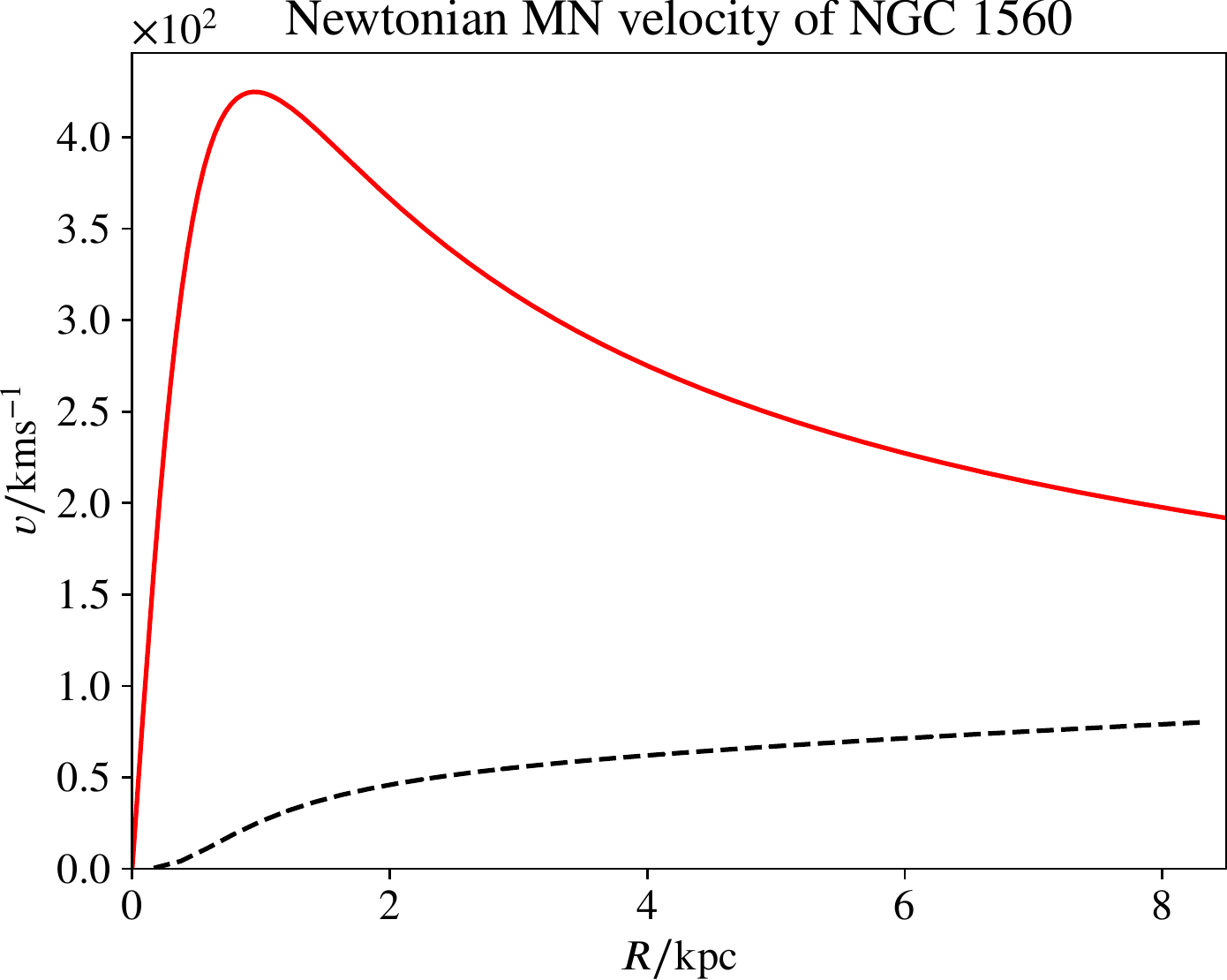}
\caption{The conventional Newtonian rotation curve (red) for NGC 1560
  assuming a MN profile with the best-fit values of the parameters 
$a$, $b$ and $M$ from~\cite{Ludwig2021}, together with the exact rotation curve
  including gravitomagnetic effects (black), already shown
  in Figure~\ref{fig:ngc-1560-rot-curves}.}
\label{fig:conv-vs-rot-curves}
\end{center}
\end{figure}

Nonetheless, let us assume the best-fit value of $M$ to calculate also
the rotation curve that one would obtain in the absence
gravitomagnetic effects, i.e.\ $\psi = 0$, and the galaxy is
completely static and supported just by usual pressure forces. In this
case, the rotational velocity of a test particle is merely
$\sqrt{R\partial\Phi/\partial R}$ and one obtains the red curve in
Figure~\ref{fig:conv-vs-rot-curves}, which we plot alongside the exact
rotation curve (in black) from Figure~\ref{fig:ngc-1560-rot-curves},
which includes gravitomagnetic effects.
Figure~\ref{fig:conv-vs-rot-curves} matches very well with Figure~2 in
\cite{Ludwig2021}, but is worthy of further comment. First, we note that the
conventional rotation curve peaks at velocities around $420 \kms$
(readopting SI units for the moment); this is much higher than one
would expect for what is meant to be a dwarf galaxy.  Second, and more
important, we see that the effects of gravitomagnetism here are to
{\it suppress} the rotational velocity of test particles, not {\it
  enhance} them. Thus one requires a great deal more matter present in
the case with gravitomagnetic effects than that without, in order to
explain a given rotation curve level. Gravitomagnetic effects serve
here to explain only aspects of the {\it shape} of rotation curves
(here a gradually rising one), but absolutely not whether one requires
more matter than appears visible; in other words, it makes the missing
matter problem {\it worse}.

Before moving on to discuss the issue of gravitomagnetic vertical
support (or the lack thereof) in the next subsection, it is worth
noting some further aspects of the shape of the rotation curves
derived above. Although the rotation curves obtained using either
(\ref{eqn:radial-chi-eqn}) or the analytic approximation
(\ref{eqn:anal-chi}) appear to fit the rotation curve data for NGC
1560 shown in Figure~1 of~\cite{Ludwig2021} in a pleasing way, this disguises the
problem that the shape of these rotation curves changes considerably
with just small changes in the $a$ and $b$ parameters.

\begin{figure}
\begin{center}
\includegraphics[width=0.93\linewidth]{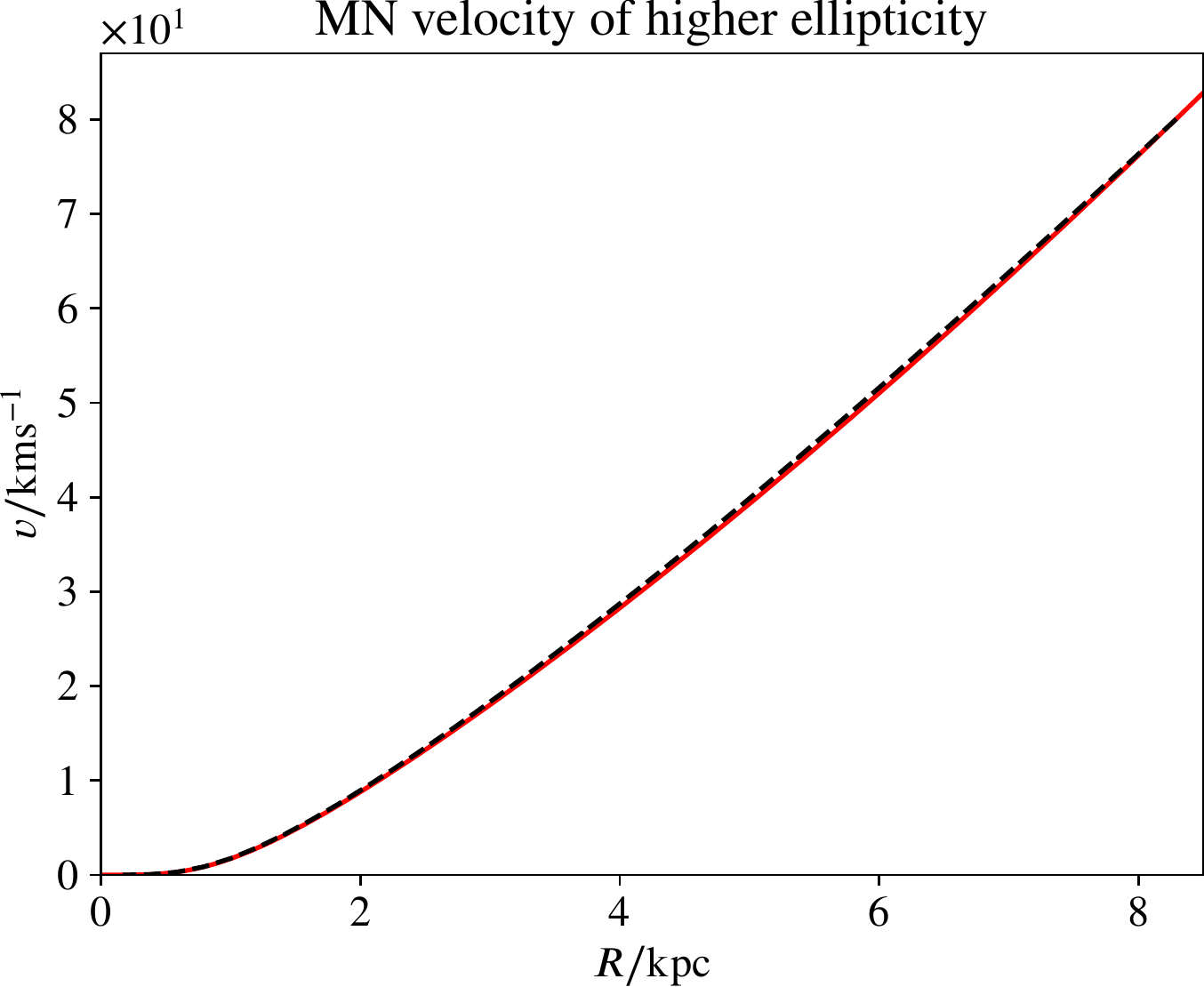}
\caption{Same as Figure~\ref{fig:ngc-1560-rot-curves}, but for a
  higher ellipticity case, with $a=\SI{0.7}{\kiloparsec}$ and
  $b=\SI{0.3}{\kiloparsec}$. \label{fig:ludwig-phi-he}}
\end{center}
\end{figure}

Observations of NGC 1560 in the visible show it to be considerably
more `elliptical' than the ratio $a:b = 0.373:0.300$ indicates, with a
ratio of $\sim 0.7:0.3$ seeming much more appropriate.  From the
analytical expression (\ref{eqn:anal-chi}), however, one can see that
this will cause a problem, since the shape of the predicted rotation
curve will scale as $v \propto R^{1.33}$ at large $R$, and so it will
be concave rather than convex towards the $R$ axis. Indeed, this will
clearly occur for any ratio $a:b > 2:1$. No known rotation curves have
this shape (concave rather than convex over their whole range), and so
this model will be incapable of accommodating galaxies with
ellipticities beyond this ratio. That this is not an artefact of our
analytical approximation is illustrated in
Figure~\ref{fig:ludwig-phi-he},
which is the equivalent of the rotation curves plot in
Fig.~\ref{fig:ngc-1560-rot-curves}, but for $a$ and $b$ values of 0.7
and $\SI{0.3}{\kiloparsec}$, and using the same mass $M$. One
 sees that the red curve (analytical approximation) closely follows
the black curve (exact numerical integration), and hence the insights
that the analytic approximation (\ref{eqn:anal-chi}) provides
for what occurs at higher $a:b$ ratios are indeed borne out in the exact
integration.

\subsection{Gravitomagnetic vertical support}\label{GravitomagneticVerticalSupport}

As our final point in this section we now discuss further the assumption that all
vertical support for dynamical equilibrium is provided by
gravitomagnetic rotational effects, which in our opinion is the key
issue with the modelling approach outlined in
Section~\ref{sec:gemmodel}, and applies irrespective of the assumed
density profile of the galaxy.  As above, however, we will illustrate
our findings for the MN profile, since it can again be treated almost
entirely analytically.

\begin{figure}
\begin{center}
\includegraphics[width=0.49\textwidth]{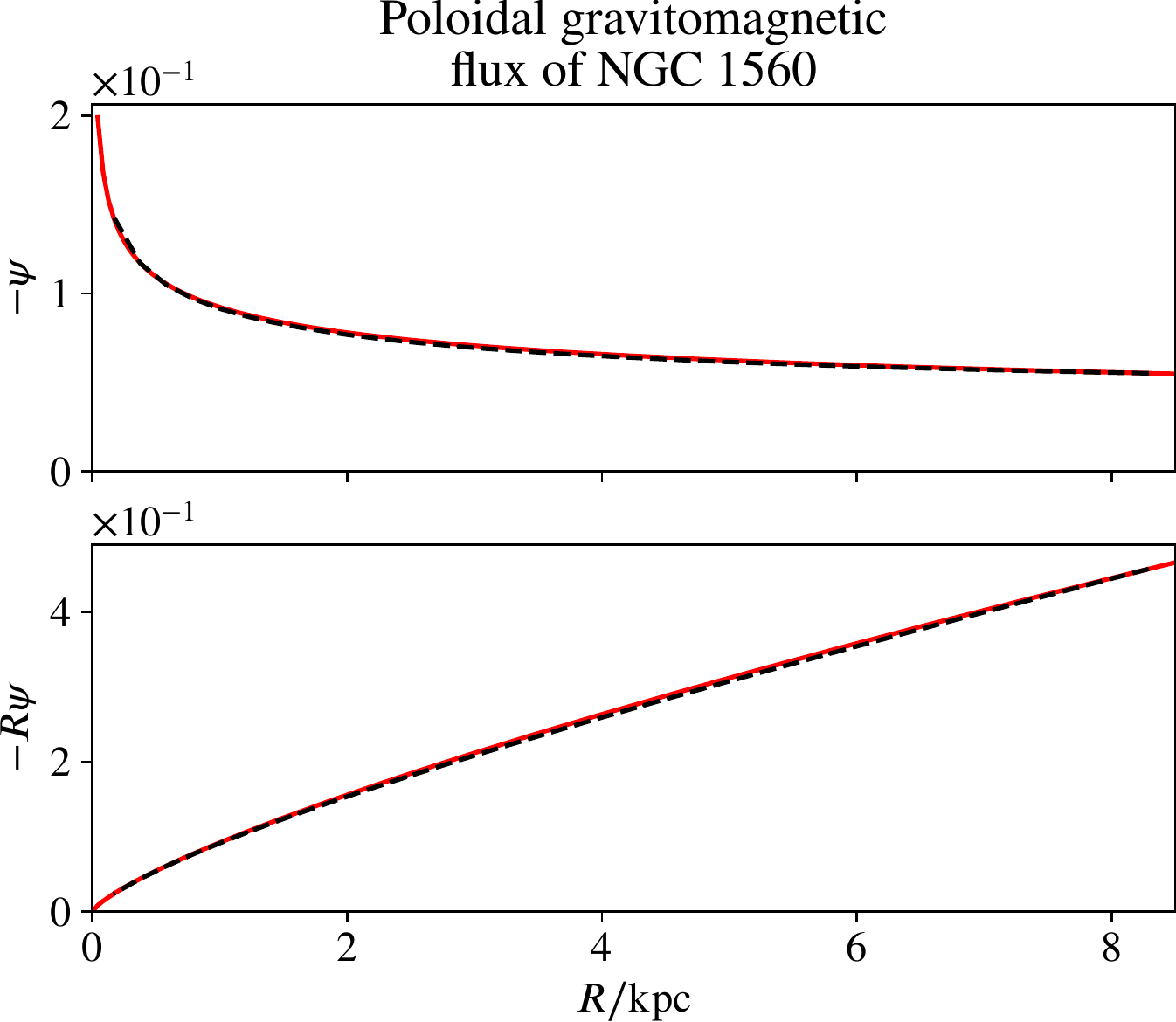}
\caption{Top: the function $-\psi$ versus $R$ in $\SI{}{\kiloparsec}$ using the
  parameters derived for NGC 1560 in~\cite{Ludwig2021}. Bottom: the function
  $-R\psi$, to indicate better the behaviour near the origin. In each
  case the black curve is the result of an exact numerical
  integration, and the red curve shows the analytic approximation
  (\ref{eqn:anal-Y}). \label{fig:ludwig-Ys}}
\end{center}
\end{figure}

In particular, we will show that in order to provide the vertical
support necessary, $\psi$ has to become infinite at the origin, and
have extremely large values near to it. To substantiate this, plus
gain some insight into what is happening analytically, we again take a
`dual track' approach in which we carry out exact numerical
integrations, as well as develop an analytical approximation. To this
end, one can construct an exact ODE in $R$ applicable in the equatorial
plane by using radial equation of motion (\ref{eomrad}), together with
our analytical approximation for circular velocity $v$ in
(\ref{eqn:anal-chi}). One can then form an approximation to $\psi$ based
on the smallness of the coefficient $C$, which yields the very simple
approximate solution
\be
\psi = \frac{M b R}{C (b-a) R^{\frac{a}{b}}}.
\label{eqn:anal-Y}
\ee
Using the values of the parameters derived for NGC 1560 in~\cite{Ludwig2021},
this approximation is in fact even better than that for the rotation
curve in (\ref{eqn:anal-chi}), as we demonstrate in
Fig.~\ref{fig:ludwig-Ys}.
The curves for the exact numerical integration (black) and the
analytic approximation from (\ref{eqn:anal-Y}) (red) are virtually
indistinguishable. One sees that $\psi$ itself diverges towards the
origin, whereas $R\psi$ converges at the origin; this is consistent
with the ratio $a/b = 0.373/0.3$ lying between 1 and 2, and hence
according to (\ref{eqn:anal-Y}) $R\psi$ should go to zero at $R=0$,
whereas $\psi$ diverges.

\begin{figure}
\begin{center}
\includegraphics[width=0.99\linewidth]{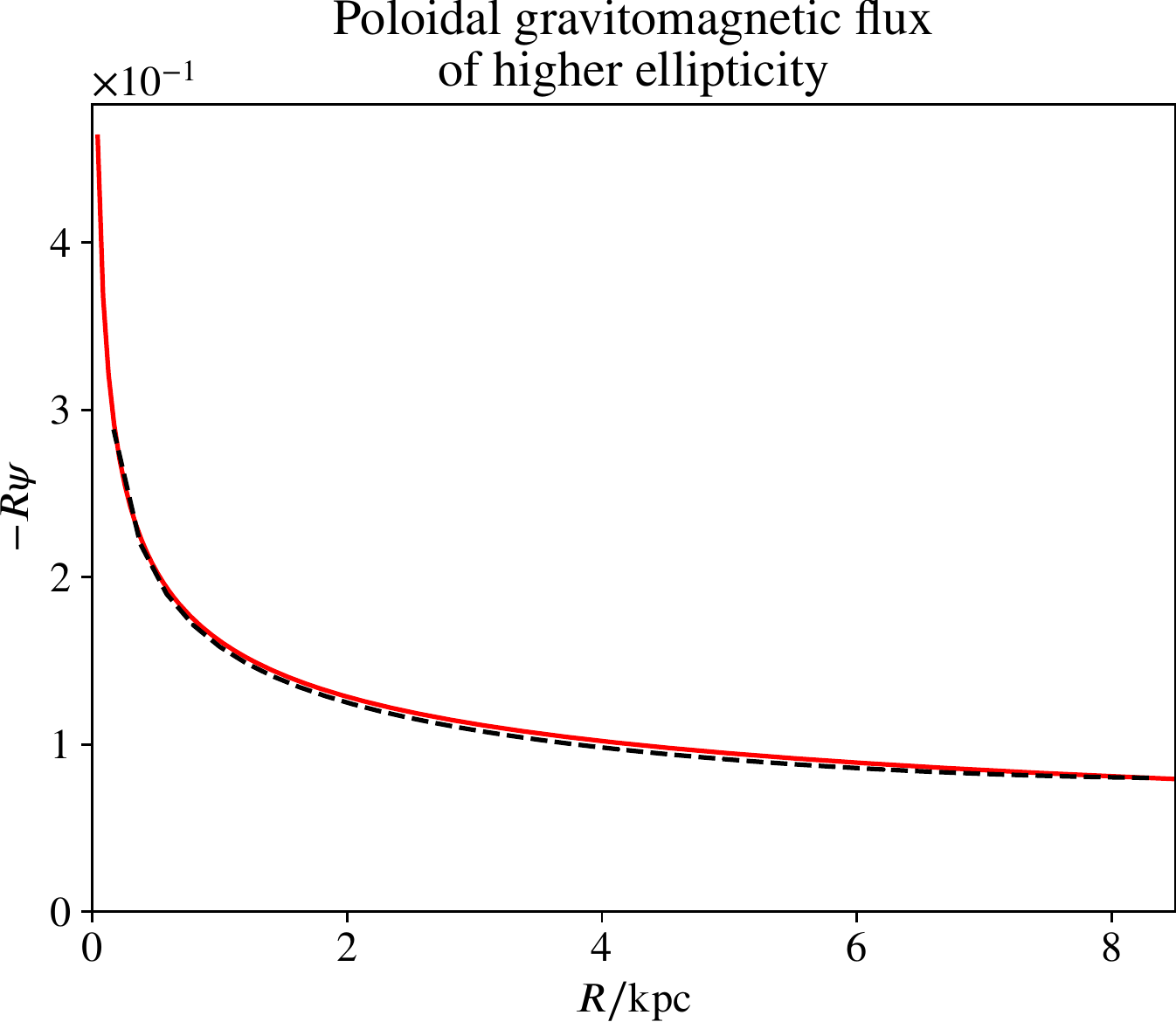}
\caption{Same as the Figure~\ref{fig:ludwig-Ys} (bottom), but for a
  higher ellipticity case, with $a/b =
  0.7/0.3$. \label{fig:ludwig-Ys-at-he-with-r-mult}}
\end{center}
\end{figure}

By comparison, in Fig.~\ref{fig:ludwig-Ys-at-he-with-r-mult}
we show $R\psi$ for the higher ellipticity case considered above,
i.e.\ $a/b=0.7/0.3$. We have plotted only $R\psi$ here since even this
diverges, as to be expected from (\ref{eqn:anal-Y}) with $a/b>2$. We
also note that in all of these plots of $\psi$ the values involved are
$\mathcal{O}(1)$ or perhaps $\mathcal{O}(10^{-1})$, which is roughly
$10^{8-9}$ larger than expected to be generated by GEM effects,
according to the orders of magnitude analysis given earlier.

This effect must originate from the unwitting inclusion of {\it free-space}
solutions of the Poisson-like equation (\ref{poissona}) that
determines $\psi$, i.e.\ solutions for which the source term on the
RHS, which would normally generate $\psi$, are set to zero. One can
introduce arbitrary amounts of such homogeneous solutions to any solution
of the inhomogeneous equation. However, the penalty is of course that
any such solution has to add in singularities at either infinity or
the origin. If this were not the case, one would be free to add
homogeneous solutions of arbitrary amplitude to, for example, the
Poisson equation for the gravitational field around the Sun or Earth,
meaning one would lose the ability to predict the force of gravity
based on the mass of an object. Such a procedure is forbidden by the
need to exclude singularities.

Thus, having demonstrated that a singularity exists (at the origin in this
case) with the GEM approach outlined in Section IV, this should definitively
rule out the methodology as a means of explaining flat galaxy rotation curves
without dark matter. It might be argued that a `get-out’ might exist since most
galaxies already contain a singularity near their centres in the form of
supermassive black holes.  However, such a model would require separate
computations that we have not seen carried out as yet to establish it, and a
priori seems contrived. Finally, although we have not gone into it here, one
finds further that a singularity can exist even if $\psi$ does not diverge,
since it turns out that to have the spacetime metric obey ‘elementary flatness’~\cite{Wilson1996}, one requires not only that $\psi$ is not divergent as $R$ approaches
zero, but must behave as $\psi\propto R$ for small $R$. The $\psi$ functions
discussed here are far from having this property, and indeed violate this
requirement all the way up the $z$-axis, posing a further problem for this line
of approach.

\section{Homogeneous poloidal solutions}
\label{HomogeneousPoloidalSolutions}

In~\cref{GravitomagneticVerticalSupport} we alluded to the unwitting inclusion in~\cite{Ludwig2021} of homogeneous solutions to the Poisson-like equation~\cref{poissona}, which can seemingly facilitate large and interesting departures from the Newtonian rotation formula. Even more recently in fact, an attempt has been made in~\cite{Astesiano:2022ghr} to capitalise directly on these solutions in an effort to bring about the same effect. In this final section, we demonstrate that the homogeneous solution approach is not viable. 

\subsection{No prospects without thin disks}\label{NoProspectsWithoutThinDisks}

We will prefer still to consider an extended, axisymmetric source, such as that of the MN density profile in~\cref{eq:MNPhi}. In contrast, the authors of~\cite{Astesiano:2022ghr} consider only an infinitesimal, equatorial thin disk with finite surface density. In the thin disk case, the poloidal gravitomagnetic flux $\psi=\psi\left(R,z\right)$ may be completely described by a Hankel-transformed function $\tilde{\psi}=\tilde{\psi}\left(\lambda,z\right)$, where
\begin{equation}\label{ThinPsiDefinition}
	\psi\left(R,z\right)=\int_0^{\infty}\mathrm{d}\lambda'Re^{-\lambda'|z|}\tilde{\psi}\left(\lambda'\right)J_1\left(\lambda'R\right).
\end{equation}
If~\eqref{ThinPsiDefinition} holds as presented in~\cite{Astesiano:2022ghr} (and we will find in~\cref{NoProspectsWithoutSources} that it does not), then in the case of an extended density profile the linearity of the vector Poisson equation~\cref{eqn:gemsolutionsA} implies that the poloidal flux at a point may be associated with a distribution of thin disks
\begin{subequations}
	\begin{align}
		\psi\left(R,z\right)&=\int_0^\infty\mathrm{d}\lambda'R\tilde\Psi\left(\lambda',z\right)J_1\left(\lambda'R\right),\label{ExtendedPsiDefinition}\\
		\tilde{\Psi}\left(\lambda,z\right)&\equiv\int_{-\infty}^{\infty}\mathrm{d}z'e^{-\lambda|z-z'|}\tilde{\psi}\left(\lambda,z'\right).\label{ExtendedPsiTildeDefinition}
	\end{align}
\end{subequations}
By substituting~\cref{ExtendedPsiDefinition} into~\cref{eomazi} and taking an inverse Hankel transform we then find
\begin{equation}\label{ThickPoloidalGradient}
\begin{aligned}
	\frac{1}{\lambda}\frac{\partial\tilde{\Psi}\left(\lambda,z\right)}{\partial z}&=\int_0^\infty \mathrm{d}R'\frac{R'}{v\left(R',z\right)}\\
	&\ \ \ \hspace{60pt} \times\frac{\partial\Phi\left(R',z\right)}{\partial z}J_1\left(\lambda R'\right),
\end{aligned}
\end{equation}
while applying the same steps to~\cref{eomrad}, in combination with the recurrence relation for Bessel functions, yields
\begin{equation}\label{ThickPoloidal}
\begin{aligned}
	\tilde{\Psi}\left(\lambda,z\right)&=\int_0^\infty\mathrm{d}R'\Bigg[ v\left(R',z\right)\\
	&\ \ \ +\frac{R'}{v\left(R',z\right)}\frac{\partial\Phi\left(R',z\right)}{\partial R'}\Bigg]J_0\left(\lambda R'\right).
\end{aligned}
\end{equation}
In the (anyway unphysical) limit of a thin disk, inspection of~\cref{ThinPsiDefinition} suggests we may be justified in using the relation
\begin{equation}\label{ThinPoloidalGradient}
	\frac{\partial\tilde{\Psi}\left(\lambda,z\right)}{\partial z}\to -\mathrm{sgn}(z)\lambda\tilde{\Psi}\left(\lambda,z\right).
\end{equation}
Precisely~\cref{ThinPoloidalGradient} is used in~\cite{Astesiano:2022ghr} to relate the integrals in~\cref{ThickPoloidalGradient,ThickPoloidal}, and this is done effectively in the singular environment of the disk itself, at $z=0$. Given this relation of integrals, the authors then take the curious step of equating the \textit{integrands}, arriving at an apparently deterministic expression for the rotational velocity at all radii
\begin{equation}\label{AstesianoEquation}
\begin{aligned}
	v\left(R',0\right)^2&=-R'\frac{\partial\Phi\left(R',0\right)}{\partial R'}\\
	&\ \ \ -\frac{R'J_1\left(\lambda R'\right)}{J_0\left(\lambda R'\right)}\left(\frac{\partial\Phi\left(R',z\right)}{\partial z}\right)_{z=0}.
\end{aligned}
\end{equation}
In~\cref{AstesianoEquation} we retain the prime on $R'$ to remind ourselves that a dummy variable has somehow ended up on the \textit{outside} of a putatively physical equation. In~\cref{AstesianoEquation} the second term on the right hand side constitutes a correction to the Newtonian rotation curve. This correction looks appealing because it is also sourced by the gravitational potential in a strict manner: the axial gravitoelectric field strength close in to the singular plane will approach the surface density of matter in the thin disk, according to the Gaussian `pill-box' construction. This correction is tunable by a ratio of Bessel functions, in which the conjugate Hankel radius appears as a single free parameter. By tuning this parameter, the poles introduced by the Bessel coefficient can be driven off to some distant extragalactic scale. The \textit{intragalactic} rotation curve on the other hand, which then looks as though it is being computed deterministically from the surface density profile, may indeed depart from the Newtonian and become flat or rising.

Whether or not~\cref{AstesianoEquation} has any physical meaning, we can at least conclude that the mathematical steps which produced it cannot be replicated without~\cref{ThinPoloidalGradient}, i.e. the construction of~\cite{Astesiano:2022ghr} \textit{requires} a singular disk. In the physical case of an extended profile,~\cref{ThickPoloidalGradient,ThickPoloidal} can only be related by differentiating under the integral sign of~\cref{ThickPoloidal}. If we then repeat the remarkable step of equating the integrands, the closest we can get to~\cref{AstesianoEquation} is the following
\begin{equation}\label{GeneralAstesianoEquation}
\begin{aligned}
	\frac{\lambda R' J_1\left(\lambda R'\right)}{v\left(R',z\right)J_0\left(\lambda R'\right)}\frac{\partial\Phi\left(R',z\right)}{\partial z}&=\frac{\partial}{\partial z}\Bigg[ v\left(R',z\right)\\
	&+\frac{R'}{v\left(R',z\right)}\frac{\partial \Phi\left(R',z\right)}{\partial R'}\Bigg].
\end{aligned}
\end{equation}
In common with~\cref{AstesianoEquation}, the true relation~\cref{GeneralAstesianoEquation} contains a deterministic correction, relative the Newtonian rotational velocity prediction, which is somewhere singular and freely tuned by the conjugate Hankel radius. However the implications of the new relation are fundamentally different: at every (dummy) radius $R'$ the velocity is determined by an ODE in the axial $z$ direction. This ODE requires some initial data for each (dummy) $R'$, which might as well be provided by some \textit{user-defined} rotation curve $v(R',0)$ in the (dummy) equatorial plane. Thus,~\cref{GeneralAstesianoEquation} requires the equatorial rotation curve as an input, and does not supply it as an output. 

\begin{figure*}
\begin{center}
\includegraphics[width=\textwidth]{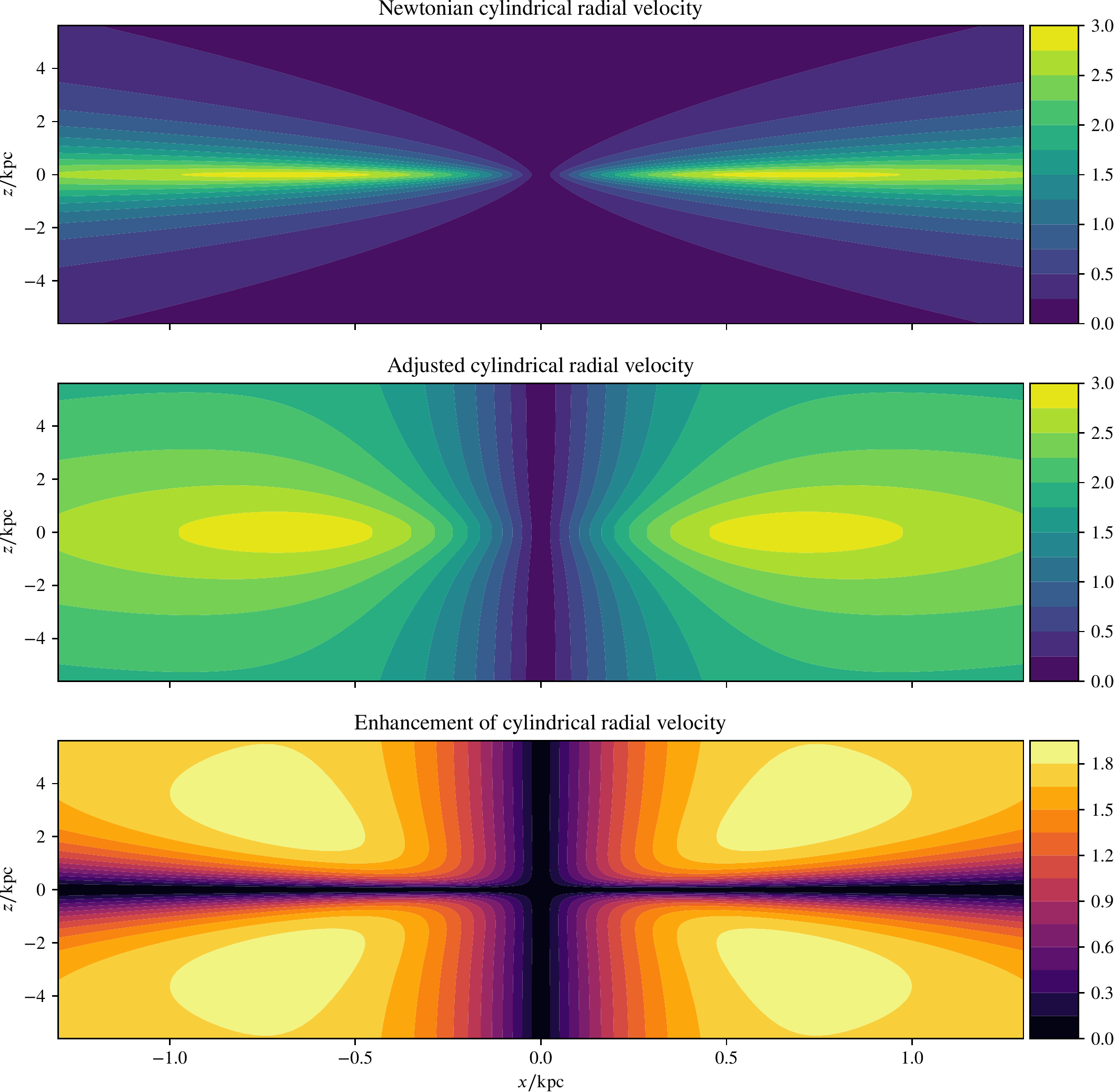}
	\caption{\label{Astesiano} Non-equatorial enhancement of the rotational velocity of the MN profile obtained by generalising the approach of~\cite{Astesiano:2022ghr} to extended sources. The Newtonian velocity is shown in the top frame for the MN potential in~\cref{eq:MNPhi}, as plotted in \cref{fig:ngc1560-contours} for parameters associated with NGC 1560 in~\cite{Ludwig2021}. The middle frame shows the enhancement in the case where the equatorial rotational velocity (initial data) is identical to the Newtonian, and~\cref{GeneralAstesianoEquation} allows this to be propagated axially using the potential. In the lower frame, the difference between enhanced and Newtonian velocities is shown, indicating a substantial velocity increase in annular zones above and below the galactic plane. The inverse Hankel radius in this case is $\lambda=\SI{1e-10}{\per\kiloparsec}$, so that the pole introduced by the first zero of the Bessel function is expelled from the observable Universe. We show that these effects result from the misuse of homogeneous solutions for the poloidal gravitomagnetic flux.}
\end{center}
\end{figure*}

If some initial data $v(R',0)$ is chosen,~\cref{GeneralAstesianoEquation} can propagate the rotational velocity axially above and below the equatorial plane, depending strictly on the gravitoelectric potential $\Phi(R',z)$ and the tunable Hankel radius $\lambda$. Because~\cref{GeneralAstesianoEquation} modifies the axial derivative of the Newtonian expression, we can still depart from the Newtonian rotational velocity above and below the equatorial plane even if we use the Newtonian expression for $v(R',0)$. This is illustrated in~\cref{Astesiano}, where the MN profile associated with NGC 1560 is used to propagate~\cref{GeneralAstesianoEquation} using precisely the Newtonian rotational velocity of that profile as initial data. With other initial data, doubtless even more interesting effects may be produced by~\cref{GeneralAstesianoEquation}: there are apparently as many possibilities as there are functions on the positive real line, and this is not the kind of situation we expect to encounter in a well posed theory of gravity such as GR. We will now clarify in~\cref{NoProspectsWithoutSources} why the construction underpinning~\cite{Astesiano:2022ghr} and our corollary in~\cref{GeneralAstesianoEquation} do not --- and can never --- arise in nature.

\subsection{No prospects without sources}\label{NoProspectsWithoutSources}

The authors of~\cite{Astesiano:2022ghr} attempt to make a distinction between what happens in determining the potential from the matter density distribution, and how the poloidal gravitomagnetic field is determined (or not) by the matter flows. In particular for the first case (density) they correctly say that our equation~\cref{poissonphi} `completely fixes the value of the Newtonian potential everywhere', whereas for the poloidal gravitomagnetic field $\psi$ in~\cref{ThinPsiDefinition}, there is meant to be a freedom in adding in homogeneous solutions of the equation which determines it. Since we do not agree with this distinction, we will start with the case of how the potential is uniquely determined by $\rho$ and then show how exactly the same procedure applied to $\psi$ again leads to unique solutions, to which we cannot add in extra homogeneous components. Despite these problems, the Hankel transform approach used by~\cite{Astesiano:2022ghr} is useful since it enables us to explicitly find the homogeneous solutions in question explicitly, and thereby show they are inadmissible.

In order to make sure that we do not introduce unnecessary singularities, we will work not with the `thin disk' approximation used by~\cite{Astesiano:2022ghr}, but continue from~\cref{NoProspectsWithoutThinDisks} with a continuous and differentiable distribution of matter, which we can call a thick disk. The MN profile used earlier would be a good example of what we have in mind here. After obtaining the results we will look at the thin disk limit, and show --- unlike in the previous analysis in~\cref{NoProspectsWithoutThinDisks} --- that it behaves in exactly the same way as found here for the thick disk. We take this as an indicator that we are finally connecting with the correct physics, and that our results are equally applicable to the case treated in~\cite{Astesiano:2022ghr}.

We thus start with the equivalent of~\cref{ExtendedPsiDefinition}, but for potential rather than the poloidal field, and write 
\be  
-\Phi(R,z) = \int_0^\infty f\left(\lambda',z\right) J_0\left(\lambda' R\right) \lam \mathrm{d}\lambda'.
\label{eqn:a1-vi-form}
\ee
The function $f(\lambda,z)$ is thus the Hankel transform, in the $R$ direction, of minus the potential.

Note particularly that, contrary to what is done in~\cref{ExtendedPsiTildeDefinition}, at this stage we are \textit{not} going to assume a particular form for $f(R,\lambda)$. This is because we will be able to \textit{deduce} the equivalent form for $f$ from the equations themselves, which is an interesting feature of the approach here.

We now insert (\ref{eqn:a1-vi-form}) into the Poisson equation for $\Phi$, obtaining 
\be  
\int_0^{\infty} \left(\frac{\partial^2 f}{\partial z^2} -\lambda'^2 f\right) J_0\left(\lambda' R\right) \lam' \mathrm{d}\lambda' =-4\pi\rho(R,z).
\ee
Taking the inverse Hankel transform of each side then yields
\be  
\frac{\partial^2 f}{\partial z^2} -\lambda^2 f=-4\pi \int_0^\infty \rho\left(R',z\right) J_0\left(\lambda R'\right) R' \mathrm{d}R'.
\label{eqn:f-eqn}
\ee

This is a linear equation for $f$ which we can solve by the method of \textit{variation of parameters}. In this technique, if we know solutions of the homogeneous equation for $f$ we can use them in constructing solutions of the inhomogeneous equation via integrations involving their product with the inhomogeneous part of the equation. In the current case this yields the following full solution for $f$:
\be 
\begin{aligned}
f(\lambda,z)&= F_1(\lambda) e^{-\lambda z} + F_2(\lambda) e^{\lambda z}\\
	&\ \ \ \ -\frac{2\pi}{\lambda} \Bigg(e^{\lambda z}\int_a^z e^{-\lambda z'} \tilde{\rho}\left(\lambda,z'\right) \mathrm{d}z' \\
	&\hspace{50pt}- e^{-\lambda z}\int_b^z e^{\lambda z'} \tilde{\rho}\left(\lambda,z'\right) \mathrm{d}z' \Bigg).
\label{eqn:f-res}
\end{aligned}
\ee  
In this equation $\tilde{\rho}(\lambda,z)$ is the Hankel transform of $\rho$, i.e.
\be  
\tilde{\rho}(\lambda,z) = \int_0^\infty \rho(R',z) J_0\left(\lambda R'\right) R' \mathrm{d}R',
\label{eqn:rho-tilde-def}
\ee
while the integration lower limits $a$ and $b$ are constants, and $F_1(\lambda)$ and $F_2(\lambda)$ are arbitrary functions of $\lambda$.

One can verify explicitly, by substituting (\ref{eqn:f-res}) into (\ref{eqn:f-eqn}), that this $f$ does indeed solve the intended equation. However, it now looks as though we have got a problem, since the solution involves naked factors of $e^{\lambda z}$ and $e^{-\lambda z}$. These appear multiplying $F_1$ and $F_2$, and also multiplying the integrals in $z'$. Considering e.g.\ $F_1(\lambda) e^{-\lambda z}$, this blows up as $z \rightarrow -\infty$ for any non-zero value of $F_1$. (Note the range of $\lambda$ is from 0 to $\infty$.) Thereafter there is an integration over $\lambda$ which occurs in equation (\ref{eqn:a1-vi-form}) but no subsequent integration over $z$, and hence the singularity will persist into the final answer for $\Phi$. The only way out of this is if $F_1(\lambda)$ is strictly zero, and of course the same considerations apply to for $F_2(\lambda)$. This then looks bad for the $e^{\pm \lambda z}$ multiplying the integrals, except in this case there is a `get out'. This is that the integrals are functions of $z$ as well as $\lambda$, via the upper limit of integration. In particular if we choose the lower limit of integration $b$ to be $-\infty$ then the integral will tend to zero as $z \rightarrow -\infty$, thus potentially (depending on respective rates of convergence of the integral and the outside $e^{-\lambda z}$ factor) leading to a finite answer. Similarly, in the first integral we should let $a=+\infty$, since then as $z\rightarrow\infty$ it is possible that a finite answer can be obtained here as well. 

With these values of $a$ and $b$, and setting $F_1(\lam)$ and $F_2(\lam)$ to zero, we get
\be 
\begin{aligned}
	f(\lam,z)&=\frac{2\pi}{\lam}\Bigg(\int_z^\infty e^{-\lam(z'-z)}\tilde{\rho}(\lam,z')\mathrm{d}z'\\
	& \ \ \ \ \ \ \ \ \ \ \  + \int_{-\infty}^z e^{\lam(z'-z)}\tilde{\rho}(\lam,z')\mathrm{d}z'\Bigg),
\end{aligned}
\ee
which assembles to give
\be 
f(\lam,z)=\frac{2\pi}{\lam}\int_{-\infty}^\infty e^{-\lam|z-z'|}\tilde{\rho}(\lam,z')\mathrm{d}z',
\label{eqn:our-f}
\ee
for which convergence is assured if $\tilde{\rho}$, and therefore $\rho$ itself, behaves reasonably.

This is excellent for our purposes. We have now achieved the analogue of equation~\cref{ExtendedPsiTildeDefinition}, but with the bonus that we know it is only the \textit{inhomogeneous} part of the Poisson equation, i.e.\ the density itself, that does the `sourcing'. All possible homogeneous contributions have been killed off by the requirement that there should not be explicit $e^{\pm \lambda z}$ type factors left in the final answer.

Note that if we wanted to move towards an explicit solution for $\Phi$ from this point, we could write the solution so far as the triple integral
\be  
\begin{aligned}
	-\Phi=\int_0^\infty \lam' \mathrm{d}\lam' \int_{-\infty}^\infty & \mathrm{d}z' \int_0^\infty  R' \mathrm{d}R'
\, \frac{2\pi}{\lam'} e^{-\lam'|z-z'|} \\
	&\times\rho(R',z) J_0(\lam' R') J_0(\lam' R).
\label{eqn:a1-triple-int}
\end{aligned}
\ee
This looks forbidding, but in fact we can explicitly carry out the $\lambda$ integral by using the Bessel function identity drawn attention to in the paper~\cite{cohl1999compact} by Cohl \& Tohline, specifically their equation (14), which reads, using the current variables,
\be
	\int_0^\infty \mathrm{d}\lam' e^{-\lam'|z-z'|} J_0(\lam' R') J_0(\lam' R)
	=\frac{Q_{-1/2}\left(\chi\right)}{\pi \sqrt{R R'}} .
\label{eqn:J0-J0-res}
\ee
Here $Q_{-1/2}(\chi)$ is a Legendre function of the second kind and
\be
\chi=\frac{R^2+R'^2+(z-z')^2}{2R R'}.
\label{eqn:chi-def}
\ee
Cohl \& Tohline further say that this Legendre function is related to the complete elliptic integral of the first kind, $K$, via
\be  
Q_{-1/2}(\chi) = \mu K(\mu),
\ee
where
\be  
\mu \equiv \sqrt{\frac{2}{1+\chi}}=\sqrt{\frac{4 R R^{\prime}}{\left(R+R^{\prime}\right)^2+\left(z-z^{\prime}\right)^2}}.
\label{eqn:mu-def}
\ee
At this point, inserting these results into (\ref{eqn:a1-triple-int}), we see we have recovered~\eqref{phisol}, with all factors agreeing exactly, hence we can declare this method of approach to be successful. This is of course not surprising as regards determining the potential from the density, where we are perfectly happy with the idea that adding in extra homogeneous solutions is prohibited by the boundary conditions, but we now show in~\cref{PoloidalRepeat} that exactly the same analysis leads to the same conclusion for the poloidal gravitomagnetic field.

\subsection{Repeating the analysis for the poloidal field}\label{PoloidalRepeat}

So we pick up from equation (\ref{eqn:a1-vi-form}), but this time in a version for the poloidal field $\psi$. We will, however, re-use $f$ for the Hankel transform of this field, since then many of the above relations will look almost identical. The particular version of Hankel transform which works best in terms of substituting into the gravitomagnetic equations is
\be  
\psi(R,z) = \int_0^\infty f(\lambda',z) R J_1(\lambda' R) \lam' \mathrm{d}\lambda',
\label{eqn:psi-vi-form}
\ee
where we can see the function $Rf$ is being transformed by a $J_1$. The equation we are substituting into is 
\be  
\frac{1}{r}\frac{\partial^2\psi}{\partial r^2} -\frac{1}{r^2}\frac{\partial\psi}{\partial r} + \frac{1}{r}\frac{\partial^2\psi}{\partial z^2}=-16\pi \rho v.
\ee
We now insert (\ref{eqn:psi-vi-form}) into this, obtaining 
\be  
\begin{aligned}
	\int_0^{\infty} \left(\frac{\partial^2 f}{\partial z^2} -\lambda'^2 f\right)& J_1(\lambda' R) \lam' \mathrm{d}\lambda'\\
	&=-16\pi\rho(R,z) v(R,z).
\end{aligned}
\ee
Taking an inverse Hankel transform of each side then yields
\be  
\begin{aligned}
	\frac{\partial^2 f}{\partial z^2}& -\lambda^2 f=\\
	&-16\pi \int_0^\infty \rho\left(R',z\right) v\left(R',z\right) J_1\left(\lambda R'\right) R' \mathrm{d}R'.
\label{eqn:f-eqn-psi}
\end{aligned}
\ee
Again this is a linear equation for $f$ which we can solve by the method of variation of parameters. The full solution this time is 
\be 
\begin{aligned}
f(\lambda,z)&= F_1(\lambda) e^{-\lambda z} + F_2(\lambda) e^{\lambda z}\\
	&-\frac{8\pi}{\lambda} \Bigg(e^{\lambda z}\int_a^z e^{-\lambda z'} \tilde{j}\left(\lambda,z'\right) \mathrm{d}z' \\
	&\hspace{40pt}- e^{-\lambda z}\int_b^z e^{\lambda z'} \tilde{j}\left(\lambda,z'\right) \mathrm{d}z' \Bigg),
\label{eqn:f-res-psi}
\end{aligned}
\ee  
where we have defined a `matter current' $j=\rho v$ and $\tilde{j}$ is its Hankel transform (using a $J_1$)
\be  
\tilde{j}(\lambda,z) = \int_0^\infty \rho\left(R',z\right) v\left(R',z\right) J_1\left(\lambda R'\right) R' \mathrm{d}R'
\ee

The arguments given before about what happens as $z \rightarrow \pm \infty$ go through in exactly the same way, and we can jump straight to the final answer for $f$ which is now
\be 
f(\lam,z)=\frac{8\pi}{\lam}\int_{-\infty}^\infty e^{-\lam|z-z'|}\tilde{j}\left(\lam,z'\right)\mathrm{d}z'.
\ee

We thus recover~\cref{ExtendedPsiTildeDefinition}, except now we know that the $\tilde{\psi}$ in this \textit{has} to be the transform of the inhomogeneous source $j$, and cannot contain a free homogeneous component.

In the units used above, which have $\SI{1}{\kiloparsec}$ as the unit of length, then clearly the $j$ or $\tilde{j}$ terms will be of order $10^{-9}$ and hence far too small to give the GEM effects claimed in the approach of~\cite{Ludwig2021}, or indeed the possible substantial modifications to rotation curves claimed to be allowable in~\cite{Astesiano:2022ghr}.

If we wish to progress in the same way as in the potential case to getting an explicit integral expression for $\psi$, then this will need the analogue of (\ref{eqn:J0-J0-res}) for $J_1$'s. This reads
\be
\begin{aligned}
	\int_0^\infty \mathrm{d}\lam' e^{-\lam'|z-z'|} & J_1\left(\lam' R'\right) J_1\left(\lam' R\right)\\
	&=\frac{1}{\pi \sqrt{R R'}} Q_{1/2}\left(\chi\right),
\label{eqn:J1-J1-res}
\end{aligned}
\ee
and according to equation (23) in~\cite{cohl1999compact} we have
\be  
Q_{1/2}(\chi) = \chi \mu K(\mu) -(1+\chi)\mu E(\mu),
\ee
where $\chi$ and $\mu$ are as defined earlier in equations (\ref{eqn:chi-def}) and (\ref{eqn:mu-def}) and $E$ is the complete elliptic integral of the second kind. Thus overall we will obtain
\be  
\begin{aligned}
	\psi(R,z)= &8\int_{-\infty}^\infty \mathrm{d}z' \int_0^\infty  R' \mathrm{d}R'
\, \sqrt{\frac{R}{R'}} \rho\left(R',z\right) \\
	&\times v\left(R',z\right)  \left(\chi \mu K(\mu) -(1+\chi)\mu E(\mu)\right).
\end{aligned}
\ee

\subsection{Thin disks}

Finally, we should comment on the relation to the `thin disk' approach used by~\cite{Astesiano:2022ghr}. If we assume that
\be  
\rho(R,z)=\sigma(R)\delta(z),
\ee
where $\sigma(R)$ is the surface density, and adopt the definition for the spectral function for the potential given in equation (18) of~\cite{Astesiano:2022ghr}, i.e.
\be  
\tilde{\Phi}(\lam)=2\pi\int_0^\infty R' \sigma\left(R'\right) J_0\left(R'\lam\right)\mathrm{d}R',
\ee
then our $\tilde{\rho}$ is given by (see equation (\ref{eqn:rho-tilde-def}) above):
\be  
\begin{aligned}
	\tilde{\rho}(\lam,z)& = \delta(z) \int_0^\infty \sigma\left(R'\right) J_0\left(\lam R'\right) R' \mathrm{d}R'\\
	&=\frac{1}{2\pi} \delta(z) \tilde{\Phi}(\lam).
\end{aligned}
\ee
Hence our $f(\lam,z)$ as given by equation (\ref{eqn:our-f}) is
\be
f(\lam,z) = \frac{1}{\lam}\tilde{\Phi}(\lam) e^{-\lam|z|},
\ee
and so our expression for (minus) the potential in this case is
\be  
\begin{aligned}
	-\Phi(R,z) &= \int_0^\infty f\left(\lambda',z\right) J_0\left(\lambda' R\right) \lam' \mathrm{d}\lambda'\\
	&=\int_0^\infty \tilde{\Phi}\left(\lam'\right) e^{-\lam'|z|} J_0\left(\lambda' R\right) \mathrm{d}\lambda',
\end{aligned}
\ee
which agrees with equation (16) of~\cite{Astesiano:2022ghr} up to an overall sign.

This shows that, unsurprisingly, we can reach the thin disk results of~\cite{Astesiano:2022ghr} starting from a non-singular distribution in the case of the potential and exactly the same will go through for the poloidal field, in the sense that the thin disk results, when done correctly, must show the same behaviour as the thick-disk ones, i.e.\ the behaviour is sourced only by the `matter current' and extra homogeneous solutions are not allowed.

\section{Conclusions}
\label{sec:conc}

We have investigated the recent claims in~\cite{Ludwig2021} that one
need not consider modified gravity theories to explain flat rotation
curves, such as those observed in galaxies, without the need for dark
matter, since such curves can be explained by gravitomagnetic effects
in standard linearised GR. We have also considered the related effects obtained in~\cite{Astesiano:2022ghr}, specifically substantial gravitomagnetic corrections to the rotation curve of a galactic toy-model which are put forward as possibly being impactful in galactic dynamics.

In~\cite{Ludwig2021} the convenient GEM formalism
is adopted and, somewhat unusually, a galaxy is modeled as an axisymmetric,
stationary, rotating, non-relativistic and pressureless `dust' of
stars, all of which follow circular orbits. This approach therefore
identifies the bulk velocity distribution of the galaxy with the
velocity of stars, thereby aiming to define a self-consistent
pressureless model.

The resulting system of GEM field equations for the gravitational
(gravitoelectric) potential $\Phi$ and the poloidal gravitomagnetic
flux $\psi$, together with the radial and vertical equations of
motion, are amenable to an order of magnitude analysis. Indeed, it is
straightforward to show that gravitomagnetic effects on the circular
velocity $v$ of a star are ${\cal O}(10^{-6})$ smaller than the
standard Newtonian (gravitoelectric) effects. Thus, as one might have
expected, any modification of Newtonian galaxy rotation curves must be
negligible. More importantly, we find that the assumption in the~\cite{Ludwig2021}
model that all the vertical support necessary to maintain dynamical
equilibrium arises from gravitomagnetic effects is impossible to
satisfy; if one assumes the presence only of ordinary matter, the
gravitomagnetic effects are ${\cal O}(10^{-6})$ too small to provide
this support.

The above issues are obscured when various quantities are eliminated
between the system of equations to arrive at the single key equation
for $v$ used by~\cite{Ludwig2021}. Nevertheless, to understand how~\cite{Ludwig2021} appears
to arrive at a self-consistent pressureless model for a galaxy, we
solve this key equation for $v$ in the case of a galaxy having a
MN density profile. This allows us to establish an
intuition for the results by adopting a `dual track' approach by
performing an exact numerical integration and by developing an
accurate anayltic approximation.

Adopting the derived values of the mass, $M$, and semi-major and
semi-minor axes, $a$ and $b$, obtained by~\cite{Ludwig2021} in fitting rotation
curve data for NGC 1560, we find that the resulting rotation curve
depends only very weakly on the mass $M$. Moreover, we show that for
larger values of $M$, the rotation curve becomes independent of $M$.
In any case, if one compares the rotation curve for the fitted
parameters with the corresponding standard Newtonian rotation curve,
one finds that the effects of gravitomagnetism are to suppress the
rotational velocity of test particles, not enhance them. Thus,
although the rotation curve including gravitomagnetic effects has a
shape closer to that observed, it requires more matter to be present
than in the Newtonian case in order to explain a given rotation curve
level, which excerbates the missing matter problem.

Although the predicted rotation curve for the fitted aspect ratio $a/b
= 0.373/0.3$ matches the observed one reasonably well, this aspect
ratio is somewhat smaller than what would be inferred from
observations of NGC 1560 in the visible, which is close to $a/b =
0.7/0.3 \approx 2.33$. We show, however, that for aspect ratios $a/b > 2$,
the predicted rotation curves are concave over their entire range,
which does not match observations in any galaxy. 

The most problematic issue, however, is that in order to provide the
necessary vertical support to maintain dynamical equilibrium, the
poloidal gravitomagnetic flux $\psi$ must become singular at the
origin and have extremely large values near to it. In particular, we
show that $\psi$ must be at least ${\cal O}(10^8)$ larger than
expected from gravitomagnetic effects. This must occur because
free-space solutions of the Poisson-like equation that determines
$\psi$ are being unwittingly included, but this is forbidden if one
wishes to avoid the presence of singularities. Moreover, the large
values of $\psi$ contradict the linearised treatment implicit in the
GEM formalism. Consequently, one may rule out the GEM model proposed
by~\cite{Ludwig2021} as a means of explaining flat or rising galaxy rotation curves without
the need for dark matter.

The involvement in~\cite{Ludwig2021} of free-space solutions to the
Poisson-like equation that determines $\psi$ then leads us naturally to consider~\cite{Astesiano:2022ghr} where (although the authors emphasise that more detailed analysis is needed) such solutions are deliberately employed. The fact that the methods of~\cite{Astesiano:2022ghr} lead to a dummy integration variable appearing on the outside of a putatively physical expression is already quite suggestive. In fact, when we try to faithfully generalise the proposed approach in~\cite{Astesiano:2022ghr} from the infinitesimal thin disk limit to an extended density profile, we find that the implications for galactic rotation curves are qualitatively different from those proposed in~\cite{Astesiano:2022ghr}. The orbital velocity above and below the equatorial plane is inded determined by an ODE in the axial direction, but this ODE requires initial data which may as well be taken as the rotation curve in the plane itself. Thus, the formulation is entirely non-predictive outside the thin disk limit. Far more seriously, we show conclusively in both the thin and thick disk cases that the free-space solutions on which~\cite{Astesiano:2022ghr} relies \textit{necessarily violate the gravitomagnetic boundary value problem at the equatorial plane}: they are inadmissible without a matter current there. We note that this objection is independent from the guaranteed existence of divergent regions in the solutions (which~\cite{Astesiano:2022ghr} notes may be tuned to large radii away from the galaxy). We conclude that (i) only the inhomogeneous parts of the GEM solutions may contribute to the rotation curve, and that (ii) they do so in a predictive manner, depending on the matter source currents. In the context of GEM, derived from GR without any infrared modification, we further conclude that these matter currents must after all include a substantial `dark' component to be consistent with the observed phenomena.

\begin{acknowledgements}
	WEVB is grateful for the kind hospitality of Leiden University and the Lorentz Institute, and is supported by Girton College, Cambridge.
\end{acknowledgements}

\bibliographystyle{apsrev4-1}
\bibliography{Manuscript}

\end{document}